\documentstyle[12pt,aasms4]{article}

\begin{document}

\title{Effects of Jet-like Explosion in SN 1987A}
\author{Shigehiro Nagataki}
\noindent
{Department of Physics, School of Science, the University
of Tokyo, 7-3-1 Hongo, Bunkyoku, Tokyo 113, Japan}\\

\begin{abstract}

We study the effects of jet-like explosion in SN 1987A. Calculations
of the explosive nucleosynthesis and the matter mixing in a jet-like
explosion are performed and their results are compared with the observations 
of SN 1987A. It is shown that the jet-like explosion model is favored
because the radioactive nuclei $\rm ^{44}Ti$ is produced in a
sufficient amount to explain the observed luminosity at
3600 days after the explosion. This is because the active alpha-rich freezeout
takes place behind the strong shock wave in the polar region.
It is also shown that the observed line profiles of $\rm Fe[II]$
are well reproduced by the jet-like explosion model. In particular, the
fast moving component travelling at (3000-4000) km/s is
well reproduced, which has not been reproduced by the spherical
explosion models. Moreover, we conclude that the favored degree of a
jet-like explosion to explain the tail of the light curve is
consistent with the one favored in the calculation of the matter mixing.
The concluded ratio of the velocity along to the 
polar axis relative to that in the equatorial plane at the Si/Fe
interface is $\sim 2:1$. This conclusion will give good
constraints on the calculations of the dynamics of the collapse-driven
supernova. We also found that the required amplitude for the
initial velocity fluctuations as a seed of the matter mixing is $\sim 30 \%$.
This result supports that the origin of the fluctuations is the dynamics 
of the core collapse rather than the convection in the progenitor.
The asymmetry of the observed line profiles of $\rm Fe[II]$ can be
explained when the assumption of the equatorial symmetry of the system is
removed, which can be caused by the asymmetry of the jet-like
explosion with respect to the equatorial plane.
In the case of SN 1987A, the jet on the north pole has to be stronger
than that on the south pole in order to reproduce the observed asymmetric line
profiles. Such an asymmetry may also be the origin of the pulsar kick.
When we believe some theories that cause such an asymmetric explosion, the
proto-neutron star born in SN 1987A will be moving in the southern part of
the remnant.

\end{abstract}

\section{INTRODUCTION} \label{intro}
\indent

The collapse-driven supernovae have been playing a great important
role in the history of the chemical evolution in the universe. Almost
all of astronomers believe that the universe was born once upon a time
(e.g. \cite{weinberg72}). Then the baryon number is
violated in the space (e.g. through the electroweak phase transition;
\cite{kuzmin85}). Light nuclei such as Li, Be, and B are synthesized
by the big-bang nucleosynthesis (e.g. \cite{walker91}). Finally,
heavy nuclei are synthesized in massive stars and ejected into the
space through the supernova explosions (e.g. \cite{arnett96}).

In particular, the collapse-driven supernovae have been making a great
contribution to the enrichment of heavy elements in the universe.
$\alpha$ nuclei, such as O, Si, and S are mainly synthesized in the
collapse-driven supernovae (\cite{hashimoto95}; \cite{timmes95}).
Their contribution to the enrichment of heavy elements in our galaxy
is about ten times higher than that of Type Ia supernovae
(\cite{tsujimoto95}). The chemical composition of the intracluster
medium can be also explained well by the abundance pattern of the
collapse-driven supernovae (\cite{loewenstein96}; \cite{nagataki98a}).
More advanced discussions will be presented when its chemical
composition can be determined more precisely by advanced numerical
calculations and observations. This proves the importance of studying
the phenomena of the collapse-driven supernovae.

In the present circumstances, the mechanism of the collapse-driven
supernovae can not be understood absolutely (e.g. \cite{bethe90}).
Many astronomers believe that the outline of the scenario for a progenitor
to explode has already been understood. In fact, Wilson (1985)
reported that a successful explosion can be attained when the effect
of the 'delayed' neutrino heating is taken into consideration. This is
the most promising mechanism of the delayed explosion.
However, the explosion energy inferred from his simulation ($\sim  0.4 
\times  10^{51}$ ergs) seems to be smaller than the observed one in
SN 1987A ($\sim 1 \times  10^{51}$ ergs; \cite{woosley88a};
\cite{shigeyama88}).
Generally speaking, it is said that the roles of many important
phenomena and processes other than the neutrino heating have to
be investigated in order to obtain a more realistic explosion model.
For example, the effects of convection and the neutrino opacities that
regulate the driving neutrino luminosities are investigated in order
to enhance the explosion energy in their models (\cite{herant94};
\cite{bethe95}; \cite{burrows95}; \cite{janka96}).

Other than the effects mentioned above, some researchers investigate
the effects of rotation of a progenitor (\cite{leblanc70}; \cite{muller80};
\cite{tohline80}; \cite{muller81}; \cite{bodenheimer83}; \cite{symbalisty84};
\cite{mochmeyer89}; \cite{finn90}; \cite{yamada94}).
They reported that the system becomes globally
asymmetric by means of its effects. In particular, some researchers
reported that a strong shock wave propagates along to the rotational
axis when an axial symmetry of the system is assumed. 
Yamada \& Sato (1994) showed clearly that a shock wave becomes
jet-like with an increase of the angular momentum in the iron core.
The qualitative explanation is
given as follows: the iron core in the equatorial plane can not
sufficiently collapse due to the centrifugal force. As a result, the
gravitational energy can not be released well and the strength of the
shock wave is reduced. On the other hand, the iron core around the
polar region collapses sufficiently and generates a strong shock wave.
Moreover, Shimizu, Yamada, \& Sato (1994) showed that neutrino heating 
from an oblate proto-neutron star, whose form is the consequence of
rotation (\cite{tohline80}; \cite{muller81}), also helps to cause a
jet-like explosion. An magnetized rotating proto-neutron
star can also generate a jet-like explosion because of the magnetic buoyancy
(\cite{leblanc70}; \cite{symbalisty84}).

The existence of asymmetry in a collapse-driven supernova is also
supported by the observations (e.g. \cite{burrows98}). For example,
some observations of SN
1987A suggest the asymmetry of the explosion. The clearest is the
speckle image of the expanding envelope with high angular resolution
(\cite{papaliolis89}), in which an oblate shape with
an axis ratio of $\sim 1.2 - 1.5 $ was shown. Similar results were also
obtained from the measurement of the linear polarization of the scattered
light from the envelope (\cite{cropper88}). It is noted that
no net linear polarization is induced from a spherically symmetric
scattering surface. Assuming that its shape is an oblate or prolate
spheroid, one finds that the observed linear polarization corresponds
to an axis ratio of $\sim 1.2$ for SN 1987A.
Moreover, the direction of the polarization is also found to be close to
that of the elongation of the SN debris (\cite{nisenson99}).
There is also growing evidence that the polarization is a rather common
phenomenon among the collapse-driven supernovae (\cite{mendez88};
\cite{wang96}).

However, it is noted that the propagation of a jet-like shock wave
in the only iron core is solved in the numerical calculations
mentioned above. This is because the most difficult problem with the
supernova dynamics is the penetration of the shock wave through the iron 
core, where the photo-dissociation of the nuclei occurs and much of
the thermal energy is wasted for the dissociation.
It is also noted that much more CPU time is necessary to follow the
whole explosion. Moreover, as stated above, almost all of simulations
concerned with the dynamics of the explosion resulted in failure.
So, when a numerical simulation is performed for the phenomena that occur
in the outer layers, such as the explosive nucleosynthesis, an initial
shock wave is assumed at the Si/Fe interface and its propagation is
followed.
It is also noted that a shock wave has been assumed to be spherically
symmetric in almost all of precise calculations concerned with
explosive nucleosynthesis in these layers (\cite{hashimoto95};
\cite{woosley95}; \cite{thielemann96}). So, we want to emphasize it
important to calculate
the explosive nucleosynthesis assuming a jet-like explosion and to
compare its results with the observations. If the jet-like explosion
model can explain the observations better than the spherical ones, it
means that a jet-like explosion is supported by the observations.
At the same time, if a proper degree of the jet-like explosion can be
determined from the observations, it means that we can make a
constraint on the degree by the observations. It will be a very
important information on the dynamics of the collapse-driven
supernovae.

There is only one calculation concerned with the explosive
nucleosynthesis behind a jet-like shock wave (\cite{nagataki97}).
It is reported that the radioactive element, $\rm ^{44}Ti$, can be
synthesized more in a jet-like explosion model because the active
alpha-rich freezeout takes place behind the strong shock wave. This
result was compared with the observations of SN 1987A,
which gives various precise data because of its
vicinity and youth. They found that $\rm ^{44}Ti$ in the jet-like
explosion model is synthesized
in a sufficient amount to explain its bolometric luminosity around
$\sim 1400$ days after the explosion. Recently, its luminosity at
$\sim 3600$ days after the explosion was reported (\cite{suntzeff99}).
Since the half-life of $\rm ^{44}Ti$ is longer than those of other main
radioactive nuclei such as $\rm ^{56}Co$ and $\rm ^{57}Co$, the recent 
data is more useful to extract the contribution of $\rm ^{44}Ti$ and
determine its amount. In this paper, the proper degree of the jet-like
explosion in SN 1987A is estimated using this observation. At the same 
time, its degree can be estimated using an independent observation,
the line profile of Fe[II] (\cite{nagataki98b}). We will present its
further discussions in this paper. Finally, we will conclude the
proper degree of jet-like explosion in SN 1987A in order to explain
these observations. The concluded ratio of the velocity along to the
polar axis relative to that in the equatorial plane at the Si/Fe
interface is $\sim 2:1$. Discussions about the meaning of
this conclusion are also presented.

In section~\ref{model}, we show our method of calculation. We consider 
the problem with the explosive nucleosynthesis in section~\ref{light}.
Matter mixing is discussed in section~\ref{mixing}. Summary is
presented in section~\ref{summary}.

\section{METHOD OF CALCULATION} \label{model}
\indent

In this section, we present the method of the simulations in this
study. It is true that there are some assumptions and simplifications
in our simulations. There are two reasons for such treatments. One is
that the supernova dynamics can not be followed from the onset of the
gravitational collapse, as mentioned in section~\ref{intro}. We have
to start a simulation assuming an initial shock wave at the Si/Fe
interface. The other is that this is the first step for the
calculation of the explosive nucleosynthesis solving a
multi-dimensional hydrodynamics in a collapse-driven supernova.
We note that there has been no such a
simulation before Nagataki et al. (1997). We have to improve the
accuracy of the simulation step by step. We explain what is assumed
and simplified in the following subsections.

\subsection{Hydrodynamics} \label{hydro}
\subsubsection{The Scheme} \label{scheme}
\indent

We performed 2-dimensional hydrodynamic calculations. The calculated
region corresponds to a quarter part of the meridian plane under the
assumption of axisymmetry and equatorial symmetry. The number of
meshes is $300 \times 10$ (300 in the radial direction and 10 in the
angular one) for the calculation of explosive nucleosynthesis and
$2000 \times 100$ for the matter mixing. Higher resolution is needed for 
the matter mixing since growth of a small initial fluctuation has to
be followed.
We did not perform 3-dimensional hydrodynamic
calculations but 2-dimensional ones to save CPU time and memory size
of the supercomputer. In fact, there is no 3-dimensional calculation
with high resolution in the world for the matter mixing in a supernova
ejecta. Much more CPU time and memory size beyond the
capacity of presently available supercomputers are necessary in order
to obtain quantitative results with 3-dimensional calculations.
However, the propagation of a jet-like shock wave along with the
polar axis, where the system is calculated 3-dimensionally even in a
2-dimensional calculation, is performed in this study. So dimensional
effects seem to be little.
The Roe scheme,
which solves Riemann's problem with high accuracy and little CPU time,
is used for the calculations (\cite{roe81}; \cite{yamada94}). So it is
a good scheme for calculating the propagation of a shock wave in a
star. However, an Eulerian coordinate is used for the
multi-dimensional calculations. That is why we use the test particle
method (\cite{nagataki97})
in order to obtain the informations on the time evolution of the
physical quanta on the frame comoving with the matter, which are used
for the calculations of the explosive nucleosynthesis.
We briefly explain the test particle method. Test particles are
scattered in a star and at rest at first. These move with the local
velocity at their own positions after the passage of the shock wave. 
At each time step, temperature and density that each test particle
experiences are preserved. This is the test particle method we use.
See Nagataki et al. (1997) and Nagataki et al. (1998b) for more details.

We comment on the assumptions and simplifications in this study.
Calculations of hydrodynamics and explosive nucleosynthesis are
performed separately, since the entropy produced during the explosive
nucleosynthesis is much smaller ($\sim$ a few $\%$) than that generated by
the shock wave.
In calculating the total yields of elements, we
assume that each test particle has its own mass determined
from their initial distribution so that their sum
becomes the mass of the layers where these are scattered, and also assume
that the nucleosynthesis occurs uniformly in each mass element.
These assumptions will be justified since the movement of the test
particles is not chaotic (i.e. distribution of the test particles at final
time still reflects the given initial condition) even in a
calculation of the matter mixing (see Figures~\ref{fig9}
and~\ref{fig10}) and
the intervals of test particles are sufficiently narrow to investigate
the contours of the chemical composition in the ejecta (see
Figures~\ref{fig3},~\ref{fig4},~\ref{fig5}, and~\ref{fig6}).

\subsubsection{Initial Conditions} \label{initial}
\indent

We comment on the initial conditions.
The initial velocity behind a shock wave is assumed to be
radial and proportional to $r \times (1+\alpha \cos(2 \theta))/
(1+ \alpha)$, where r, $\theta$, and $\alpha$ are radius, the zenith
angle, and the free parameter that determines the degree of the
jet-like explosion.
Since the ratio of the velocity in the polar region to that in the
equatorial one is $(1+\alpha)/(1-\alpha)$ : 1,
more extreme jet-like shock waves are obtained as $\alpha$ gets larger.
In this study, we take $\alpha=0$ for the spherical explosion 
and $\alpha=\frac{1}{3},\; \frac{3}{5}, \; \rm and \; \frac{7}{9}$
(these values mean that the ratios of the velocity are 2:1, 4:1, and 8:1,
respectively) for the jet-like ones. We name these models S1, A1, A2,
and A3, respectively.
We assumed that the distribution of thermal energy is same as the velocity
distribution and that total thermal energy is equal to total kinetic one. 
The explosion energy is set to be $1 \times 10^{51} \rm
ergs$ and injected to the region of $(1.0-1.5) \times 10^{8} \rm cm$
(that is, at the Fe/Si interface).
We notice again that the form of the initial shock wave can not be known
$\it a \; priori$. That is why we take
$\alpha$ as a free parameter. Its proper value is inferred from the
comparison of calculations with the observed
luminosity and line profiles of Fe$\rm [II]$ in SN 1987A in this study.

As for the seed of the matter mixing, we perturb only velocity field
inside the shock wave when it reaches the He/H interface.
For the jet-like explosion, we introduce the perturbation when
the shock front in the polar region reaches the interface. We adopt
the monochromatic perturbations, i.e., $\delta v \; = \; \varepsilon
v(r,\theta) \cos(m \theta)$ ($m$=20; \cite{hachisu92};
\cite{yamada91}). It is noted that test particle method will be more
valid when we adopt the monochromatic perturbations rather than the random
perturbations. This is because the movement of the test particles becomes
chaotic when the random perturbations are adopted. Moreover, as the
mixing width, i.e., the length of the mushroom-like 'fingers', depends
mainly on the density structure of the presupernova model and on the
value of $\varepsilon$ rather than its mode (\cite{hachisu92}), we
think we can use this monochromatic perturbation method in order to estimate
the degree of the matter mixing and construct the line profiles of the 
heavy elements.
$0 \%$, $5 \%$, and $30 \%$ are taken for the value of $\varepsilon$.
The reason why we take $5 \%$ and $30 \%$ for the amplitudes of the
perturbation is explained in subsection~\ref{problem2}.
Models for the initial shock wave are summarized in Table~\ref{tab1}.

\placetable{tab1}

As for the progenitor of SN 1987A, Sk-69$^{\circ}$202, it is thought to
have had the mass $\sim 20 M_{\odot}$ in the main-sequence stage
(\cite{shigeyama88}; \cite{woosley88}) and had $\sim$
(6$\pm$1)$M_{\odot}$ helium core (\cite{woosley88a}).
In this study, the presupernova model obtained
from the evolution of 6 $M_{\odot}$ helium core (\cite{nomoto88})
is used for the calculation of explosive nucleosynthesis.
However, it is reported that this 6 $M_{\odot}$ model seems to be
neutron-rich in the Si-- rich layer and that $Y_e$ in the layer has to be
changed higher to suppress the overproduction of neutron-rich nuclei
(\cite{hashimoto95}). This means that the range of convective mixing
in the presupernova model is artificially changed
(\cite{hashimoto95}). This treatment is also justified when effects of 
the delayed explosion are taken into consideration
(\cite{thielemann96}). The reason is as follows. The system behind the shock
wave is photon-dominated and the temperature is almost determined as a
function of radius, irrespective of the density. So the range where the
explosive silicon burning occurs becomes wider when the matter falls toward the
central compact object before the delayed explosion occurs. When we
guess the location of the mass cut (the boundary between the ejecta
and the central compact object) by the amount of $\rm ^{56}Ni$ in the 
ejecta (this is explained in subsection~\ref{masscut}), the mass cut
is located at a more outer layer, where $Y_e$ is higher than that in the
inner layer. Making $Y_e$ in the inner Si-- rich layer higher will reflect 
this tendency effectively. In this study, we changed the value of $Y_e$
between $M = 1.637 M_{\odot}$ and the Si/Fe interface to that at $M =
1.637 M_{\odot}$ (= 0.499).
As for the
calculation of matter mixing, hydrogen envelope with 10.3 $M_{\odot}$
is attached on the helium core. This is because a mass loss of a few
solar masses is thought to have occurred before the explosion
(\cite{saio88}).

\subsection{Nuclear Reaction Network} \label{nuclear}
\indent

Since the chemical composition behind the shock wave is not in nuclear 
statistical equilibrium, the explosive nucleosynthesis has to be
calculated using the time evolution of $(\rho,T)$ and a nuclear
reaction network. $(\rho,T)$ on the frame comoving with the matter can
be obtained by means of the test particle
method (see subsubsection~\ref{scheme}). The nuclear reaction network
contains 250 species (see Table~\ref{tab2}). We add some species around
$\rm ^{44}Ti$ to Hashimoto's network that contains 242 nuclei
(\cite{hashimoto89}), since we focus mainly on its abundance in
section~\ref{light}. But, it turned out that the result was not
changed effectively by the addition.

\placetable{tab2}

\subsection{The Way of Determining the Mass Cut} \label{masscut}
\indent

In a collapse-driven supernova, there is a boundary between the
ejecta and the central compact object. This boundary is called as a
mass cut. It is a very difficult problem to determine its location
(\cite{woosley95}; \cite{hashimoto95}; \cite{thielemann96};
\cite{shigeyama98}; \cite{nakamura99}). If the dynamics of a
collapse-driven supernova can be followed from the onset of the core
collapse to the final explosion, the location of the mass cut will be
determined consistently by the hydrodynamic simulation. However, as
stated in section~\ref{intro}, such a calculation has not been
reported yet. Even if a location of the mass cut is determined by a
hydrodynamic calculation for the explosive nucleosynthesis, it will be
influenced by the initial condition, such as the initial shock wave
given at the Si/Fe interface.
Moreover, it is very difficult to determine its location
hydrodynamically since it is sensitive not only to the explosion
mechanism, but also to the presupernova structure, stellar mass, and
metallicity. In fact, it is reported that total amount of $\rm
^{56}Ni$ observed in SN 1987A is not reproduced when the location of
the mass cut is determined by the hydrodynamic calculations
(\cite{woosley95}).

There is another way to determine its location.
Among many observations of SN 1987A, the total amount of $\rm ^{56}Ni$
in the ejecta is one of the most reliable one. Moreover, $\rm ^{56}Ni$
is synthesized in the Si-- rich and the inner O-- rich layers where
the mass cut is located. So it is reasonable to
determine its location so as to contain $\sim 0.07 M_{\odot}$ $\rm
^{56}Ni$ in the ejecta (\cite{hashimoto95}). We take the same way in
this study. However, this method is simple only for a spherical
explosion model. We must extend this guiding principle for
multi-dimensional calculations as follows.
We assume that the larger total energy (internal energy plus kinetic
energy) a test particle has, the more favorably it is ejected
(\cite{shimizu93}). Under the assumption, we first calculate the total
energy of each test particle at the final stage ($\sim$ 10 s) of our
calculations for the explosive nucleosynthesis and then add up
the mass of $\rm ^{56}Ni$ in a descending order of the total energy
until the summed mass reaches $0.07  M _{\odot}$. The rest particles are
assumed to fall back to the central compact object. In this way, the
location of the mass cut is inferred. We refer to this mass cut as A7.
To check the dependence of our analysis on the form of the mass cut, 
we take another mass cut for comparison. The form of the another mass
cut is set to be spherical and determined so as to contain $0.07
M_{\odot}$ $\rm ^{56}Ni$ in the ejecta. We refer to this mass cut as S7.

\section{EXPLOSIVE NUCLEOSYNTHESIS IN SN 1987A} \label{light}
\indent

\subsection{What are the Problems?} \label{problem1}
\indent

SN 1987A in the Large Magellanic Cloud has provided us with the
most precise data to test the validity of numerical calculations of
explosive nucleosynthesis in a collapse-driven supernova. The
bolometric luminosity began to increase in a few weeks after the
explosion (\cite{catchpole87}; \cite{hamuy87}), which is attributed to
the radioactive decays of $\rm ^{56}Ni \rightarrow ^{56}Co
\rightarrow ^{56}Fe$ (half-lives of $\rm ^{56}Ni$ and $\rm ^{56}Co$
are 5.9 and 77.1 days). Then the luminosity began to
decline at a constant rate that coincides with the decay rate of $\rm
^{56}Co$. The amount of $^{56} \rm Ni$ synthesized during the
explosion is estimated to be $(0.07-0.076) M_{\odot}$ on the basis of
the luminosity study (\cite{shigeyama88}; \cite{woosley88}).

$^{57} \rm Ni$ and $^{44} \rm Ti$ are also thought to be important
heating sources to reproduce the form of the bolometric light curve.
The decline rate is slowed down at $\sim 900$ days after the explosion
(\cite{suntzeff91}), which suggests the existence of heating sources
besides $\rm ^{56}Co$.
Since the half-lives of $^{57} \rm Co$ and $^{44} \rm Ti$ are longer
than that of $^{56}\rm Co$ (half-lives of $\rm ^{57}Ni$, $\rm
^{57}Co$, and $\rm ^{44}Ti$ are 35.6 hours, 272 days, and $\sim 60$
yrs), their relative contributions to the light curve are getting
greater and greater.
In particular, $\rm ^{44}Ti$ will be the dominant heating source among the
radioactive nuclei at late time. As stated in section~\ref{intro},
the bolometric luminosity at $\sim 3600$ days after the explosion was reported
recently (\cite{suntzeff99}). It should be emphasized that the
luminosity hardly
attenuates between 1700 and 3600 days after the explosion, which is
easily explained if $\rm ^{44}Ti$ is the dominant heating source
during the period. The required amount of $\rm ^{44}Ti$ to reproduce
the luminosity is $\sim 1.5 \times 10^{-4} M_{\odot}$.
Further explanation about this amount is presented in the next subsection.
As for the amount of $\rm ^{57}Ni$, it can also be measured by the 122 
keV line from $\rm ^{57}Co$ decay (\cite{clayton74}; \cite{kurfess92}).
X-ray light curve. It is reported that the ratio $\langle \rm
^{57}Ni/^{56}Ni \rangle = \langle \rm
^{57}Co/^{56}Co \rangle  \equiv [\it X(\rm ^{57}Ni)/ \it X(\rm ^{56}Ni)]/[\it
X(\rm ^{57}Fe)/\it X(\rm ^{56}Fe)]_{\odot}$ is $1.5 \pm 0.5$
(\cite{kurfess92}).

$\rm ^{58}Ni$ is another important nucleus that is produced at the innermost
region of the ejecta and gives informations about the location of the mass cut.
From the spectroscopic observations, the ratio $\langle \rm
^{58}Ni/^{56}Ni \rangle \equiv[\it X(\rm ^{58}Ni)/ \it X(\rm ^{56}Ni)]/[\it
X(\rm ^{58}Ni)/\it X(\rm ^{56}Fe)]_{\odot}$ is estimated to be 0.7-1.0
(\cite{rank88}; \cite{witterborn89}; \cite{aitken89}; \cite{meikle89};
\cite{danziger91}).

We have to reproduce these amounts mentioned above in the numerical
calculations. We note that some excellent numerical calculations for
the explosive nucleosynthesis in SN 1987A assuming a spherical
explosion have been already reported (\cite{hashimoto95}; \cite{woosley95};
\cite{thielemann96}; \cite{nagataki97}, hereafter we call them Ha95,
WW95, TNH96, Na97, respectively). Their results are shown in
Table~\ref{tab3}. We first compare their results with the
observations.

\placetable{tab3}

It is a matter of course for Ha95, TNH96, and Na97 to be able to
reproduce the observed amount of $\rm ^{56}Ni$, because the location
of the mass cut
is determined using this observed value (see subsection~\ref{masscut}).
As for WW95, its location is determined hydrodynamically and $\rm
^{56}Ni$ is ejected more than the observed value. The calculated
amounts of $\rm ^{57}Ni$ in Ha95, TNH96, and Na97 are in the range of
the observed value. The amount in WW95 seems to be smaller than the
observed one. On the other hand, the amount of $\rm ^{58}Ni$ is well
reproduced in WW95, while the results of the other models seem to be
larger. This may reflect the neutron richness of the progenitor that
Ha95, TNH96, and Na97 used (\cite{nomoto88}; \cite{hashimoto95}, see also
subsubsection~\ref{initial}). Finally, the calculated amounts of $\rm
^{44}Ti$ in Ha95, WW95, and Na97 are smaller than the required value.
In fact, it is reported that $\rm ^{44}Ti$ cannot be produced in a
sufficient amount to explain the tail of the light curve in a wide
parameter range (\cite{woosley91}). Only TNH96 reproduces the required
value very well.

As stated above, the calculated abundances of these nuclei do not
converge among these models. This is because there are some
differences of the input physics, such as the treatment of convection
in a progenitor, nuclear reaction rates, and the way to initiate a
shock wave among these models (\cite{aufderheide91};
\cite{nagataki98c}). Unfortunately, the convergence has not been
attained yet, although some efforts for the convergence has already
started (\cite{hoffman98}). In the present circumstances, it is
necessary to make our standpoint clear. Our standpoint in this study
is as follows. We will present a consistent solution for the problems
of the explosive nucleosynthesis and matter mixing.
That is, it is shown in this study that more $\rm ^{44}Ti$ is synthesized and
ejected along with the degree of the jet-like explosion. So we can
determine the favored degree of a jet-like explosion to explain the
tail of the light curve. Next, we also show that the favored degree
for the explosive nucleosynthesis is consistent with the one favored in the
calculation of the matter mixing. So we conclude that this is a
solution for these problems. It is noted that there has been no model
that solves these problems at the same time.
This is our standpoint in this study.

\subsection{Required Amount of $\rm ^{44}Ti$ in SN 1987A}\label{required}
\indent

\begin{figure}
\epsscale{1.0}
\plotone{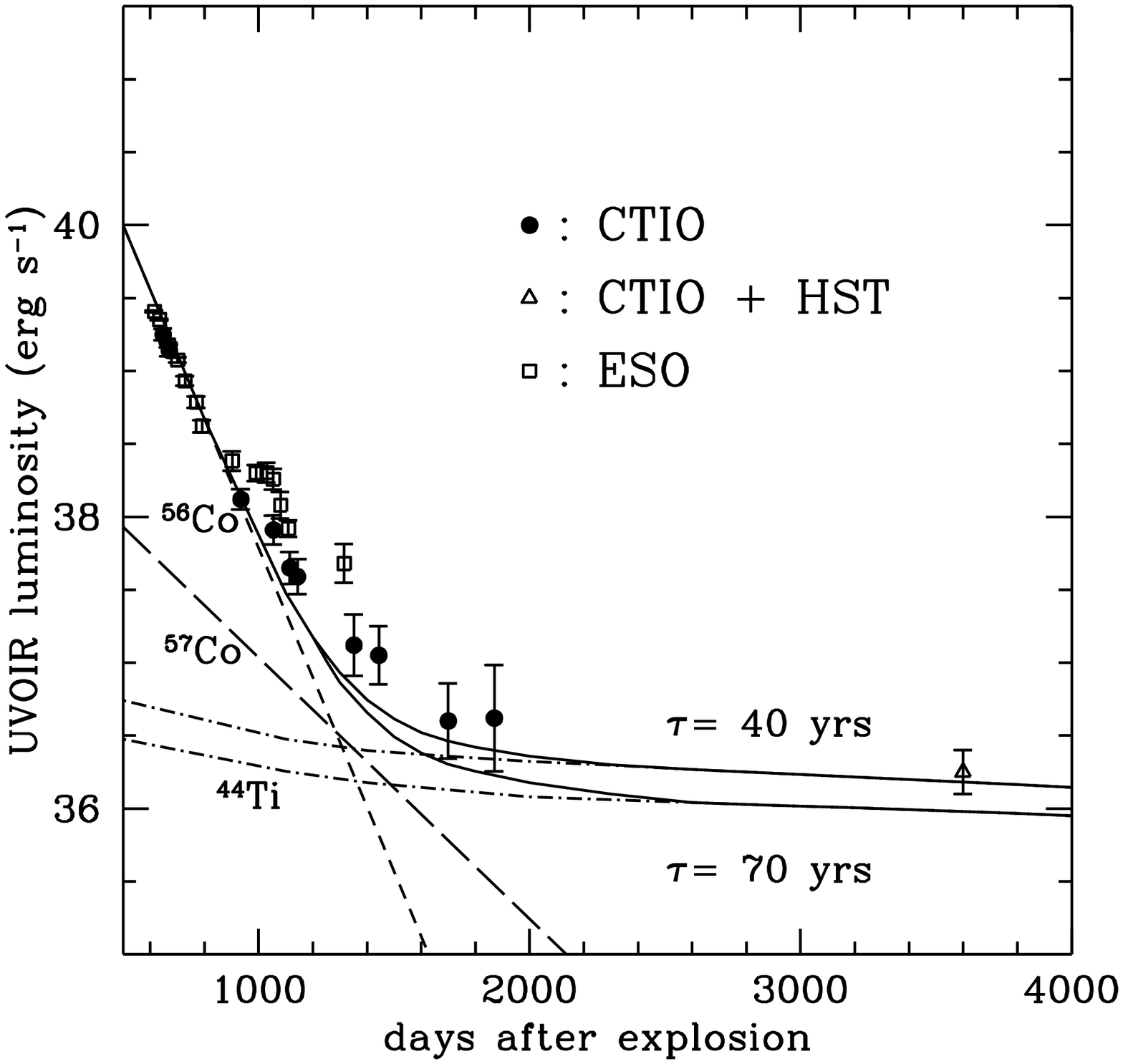}
\figcaption{
Observed bolometric (UVOIR) luminosity as a function of the days after the
explosion. Filled circles: observational data of CTIO. Open squares:
those of ESO. Triangle at $\sim$ 3600 days after the explosion is
the observed luminosity obtained by the collaboration of CTIO with
HST. Theoretical curves (Mochizuki et al. 1999a) are also
shown in figure.  
Solid curves are the total bolometric luminosities (half-life of $\rm
^{44}Ti$ is assumed to be 40 and 70 yrs). Relative contributions of
$\rm ^{56}Co$, $\rm ^{57}Co$, and $\rm ^{44}Ti$ are also shown in the
figure. The amount of $\rm ^{56}Co$ is assumed to be 0.073$M_{\odot}$.
$\langle \rm ^{57}Co/^{56}Co \rangle \equiv [\it X(\rm ^{57}Ni)/ \it
X(\rm ^{56}Ni)]/[\it X(\rm ^{57}Fe)/\it X(\rm ^{56}Fe)]_{\odot}$ is
set to be 1.7. $\langle \rm ^{44}Ti/^{56}Ni \rangle \equiv [\it X(\rm
^{44}Ti)/ \it X(\rm ^{56}Ni)]/[\it X(\rm ^{44}Ca)/\it X(\rm
^{56}Fe)]_{\odot}$ is assumed to be 1.0.
\label{fig1}}
\end{figure}

In this subsection, the required amount of $\rm ^{44}Ti$ is
explained in detail.
Observed bolometric luminosity (e.g. \cite{suntzeff90}; \cite{suntzeff91};
\cite{bouchet91a}; \cite{bouchet91b}; \cite{suntzeff92}) is shown in
Figure~\ref{fig1}.
The amount of $\rm ^{56}Ni$ is determined by the luminosity around (600
-- 900) days after the explosion. The luminosity at $\sim 3600$ days
after the explosion (\cite{suntzeff99}) is also plotted in the figure.
It is quite apparent that the decline rate becomes smaller and
smaller along with time. In particular, we can find that the
luminosity hardly attenuates between 1700 and 3600 days after the
explosion. This feature can be explained if $\rm ^{44}Ti$ becomes the
dominant isotope after $\sim$ 1700 days (\cite{kozma99}).

Before we continue to further discussion,
we must comment on the discrepancy between the observations
of CTIO (Cerro Tololo Inter-American Observatory) and ESO (European
Southern Observatory) around (1000 -- 1200) days after the
explosion. Mochizuki et al. (1998, 1999a, 1999b) obtained the required
amount of $\rm ^{44}Ti$ using the observations of CTIO and HST at
$\sim 3600$ days. In their study, they try to give an unified
explanation for the observations CTIO and HST. That is, the bolometric
luminosity evolution deduced from the observations of CTIO and HST is
reproduced in their light curve models. We use their result in this study. 
As for the observations of ESO, it
seems to be too difficult to explain such a high luminosity around
(1000 -- 1200) days because the amount of $\rm ^{57}Ni$ is limited by
the X-ray observation (\cite{kurfess92}) and numerical calculations
(\cite{hashimoto95}; \cite{woosley95}; \cite{thielemann96}). It is
also shown that such a large amount of $\rm ^{57}Ni$ to reproduce the
observations of ESO can not be synthesized in this study.

\placefigure{fig1}

Mochizuki et al. (1998, 1999a) obtained the required amount of $\rm
^{44}Ti$ as a function of its half-life using the Monte Carlo
simulation method (\cite{kumagai91}; \cite{kumagai93}).
Their theoretical light curves (\cite{mochizuki99a})
are also shown in Figure~\ref{fig1}.
Solid curves are the total bolometric luminosities (half-life of $\rm
^{44}Ti$ is assumed to be 40 and 70 yrs). Relative contributions of
$\rm ^{56}Co$, $\rm ^{57}Co$, and $\rm ^{44}Ti$ are also shown in the
figure. The amount of $\rm ^{56}Co$ is assumed to be 0.073$M_{\odot}$.
$\langle \rm ^{57}Co/^{56}Co \rangle$ is set to be 1.7. $\langle \rm
^{44}Ti/^{56}Ni \rangle \equiv [\it X(\rm ^{44}Ti)/ \it X(\rm ^{56}Ni)]/[\it
X(\rm ^{44}Ca)/\it X(\rm ^{56}Fe)]_{\odot}$ is assumed to be 1.0. We
can find $\rm ^{44}Ti$ becomes the dominant heating source among the
radioactive nuclei at late time ($\geq 1600$ days after the explosion).

The required amount of $\rm ^{44}Ti$ in SN 1987A depends on its
half-life since its decay rate means the rate of energy release from
the nuclei. This proves the importance to verify the value of its
half-life. Historically speaking, there was a large spread in the published 
values for its half-life. Its upper limit obtained by experiments is
$\tau = 66.6$ yrs (\cite{alburger90}), while its lower limit is
$\tau = 39$ yrs (\cite{meissner96}).
However, in spite of the difficulty in determining the half-life of
the long-lived nuclei, recent experiments give a converged value,
$\tau \sim 60$ yrs. For example, Ahmad et al. (1998), G$\rm
\ddot{o}$rres et al. (1998), and Norman et al. (1998) reported it to
be 59.0, 60.3, and 62 yrs, respectively. This is because the skills of the
experiments in reducing the systematic uncertainties gain (\cite{gorres98}).
Mochizuki et al. (1998, 1999a) obtained the required amount of $\rm
^{44}Ti$ as a function of its half-life using the Monte Carlo
simulation method (\cite{kumagai91}; \cite{kumagai93}).

There are other approaches to determine the amount of
$\rm ^{44}Ti$ in SN 1987A. One of them is to observe and model the
different broad bands. Since the bolometric corrections are not easy,
this method may be a better one (\cite{kozma99}).
Kozma (1999) reproduced the light curves for the B and V bands up to
4000 days for SN 1987A and reported that the required
amount of $\rm ^{44}Ti$ is in the range (1.5$\pm$1.0)$\times
10^{-4}M_{\odot}$. In her analysis, the observations from Suntzeff
\& Bouchet (1990), Suntzeff et al. (1991), and HST data
(\cite{suntzeff99}) are used. As for the half-life of $\rm ^{44}Ti$,
54 yrs is adopted in her analysis (C.Kozma, private communication).
Fortunately, the result in her analysis is consistent with the result
of Mochizuki et al. (1998, 1999a).

Still another method to estimate the amount of $\rm ^{44}Ti$ is to
observe and model individual line fluxes.
Borkowski et al. (1997) and Lundqvist et al. (1999) estimated the
required amount of $\rm ^{44}Ti$ in SN 1987A using Infrared Space
Observatory observations of the flux of Fe[II] 25.99 $\rm \mu$m line.
In particular, Lundqvist et al. (1999)
have checked several uncertainties in their model very precisely and
given a good estimation for the amount of $\rm ^{44}Ti$.
As for the half-life of $\rm ^{44}Ti$, 60.3 yrs is adopted in order to
estimate the upper limit on the amount of $\rm ^{44}Ti$ in their
analysis (\cite{lundqvist99}).
As a result, they reported that the upper limits are 1.5$\times
10^{-5}M_{\odot}$ (\cite{borkowski97}) and 1.5$\times
10^{-4}M_{\odot}$ (\cite{lundqvist99}), respectively.
Their idea to use the line flux directly will be better than that to use
the bolometric luminosity, because the bolometric corrections have to
be used to estimate a bolometric magnitude (\cite{suntzeff99}), although
no line emission, such as Fe[III] 22.93$\rm \mu$m, Fe[I] 24.05$\rm
\mu$m, and Fe[II] 25.99$\rm \mu$m, is reproduced in the
wavelength range (23 -- 27) $\rm \mu$m in their models.
In particular, the lowest upper-limit on the $\rm ^{44}Ti$ mass estimated
by Borkowski et al. (1997) from their deep ISO exposure is a puzzling
problem. Further modeling of those data are required to check the
inconsistency among their results, those of Kozma (1999)
and Mochizuki et al.(1999a).

\begin{figure}
\epsscale{1.0}
\plotone{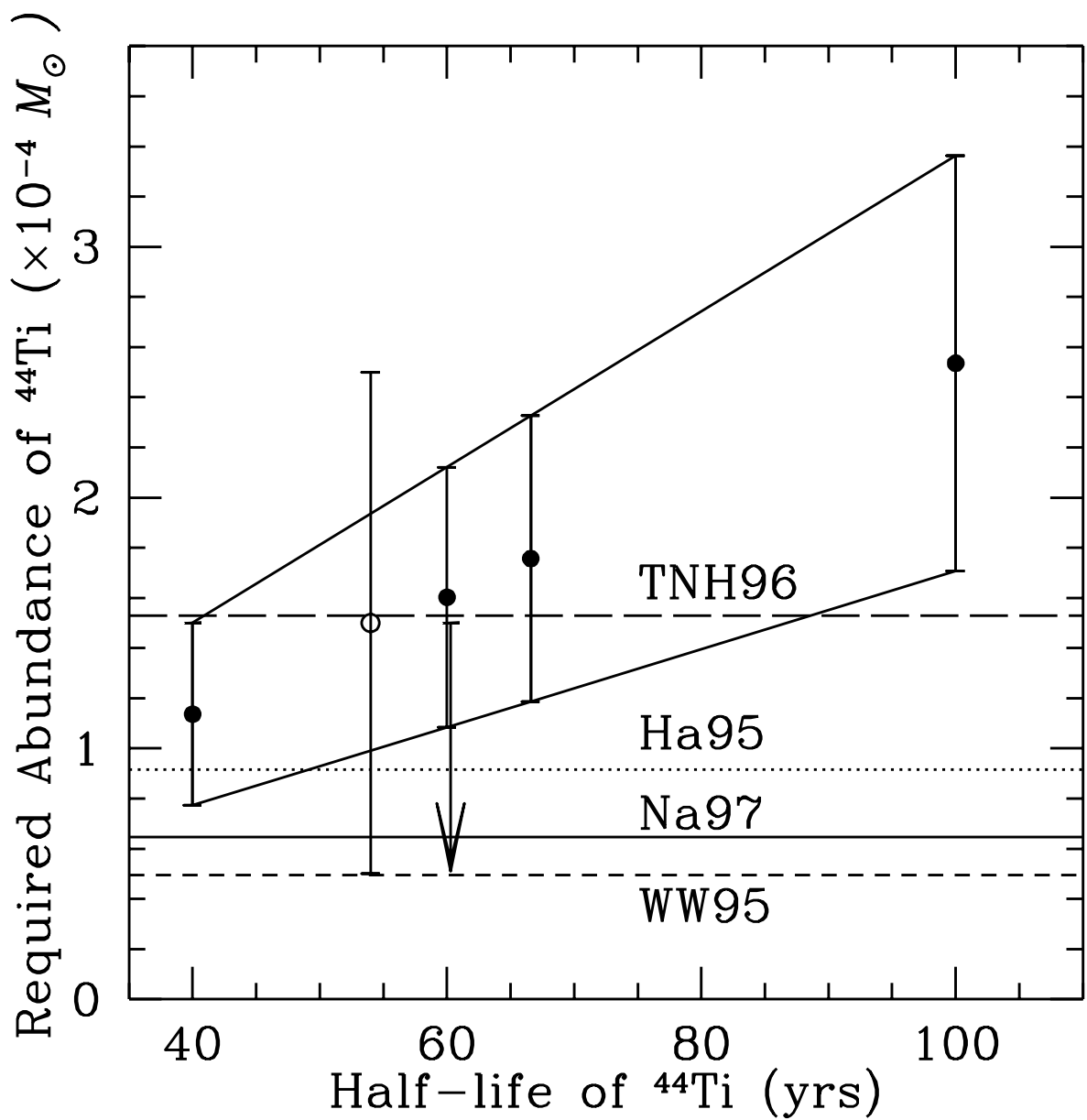}
\figcaption{
Trapezoidal region and filled circles: the amounts of $\rm ^{44}Ti$
estimated by the analysis of the bolometric light curve (Mochizuki \&
Kumagai 1998; Mochizuki et al. 1999a).
Open circle: the amount of $\rm ^{44}Ti$ estimated from the light curves for
the B and V bands (Kozma 1999). Down arrow: the upper limit on the
amount estimated by the flux of Fe[II] 25.99 $\rm
\mu$m line (Lundqvist et al. 1999). The most reliable half-life value is
$\sim$ 60 yrs (Ahmad et al. 1998; G$\rm \ddot{o}$rres et al. 1998;
Norman et al. 1998).
The upper/lower limits of the half-life obtained by the experiments
are $\tau = 66.6$ yrs (Alburger \& Harbottle 1990) and $\tau = 39$ yrs
(Meissner 1996).
Mochizuki et al. (1999a) also calculated the required amount for an
effectively longer half-life ($\sim$ 100 yrs).
Horizontal lines: results of the numerical calculations Ha95,
WW95, TNH96, and Na97. WW95 represents the model T18A in their paper.
\label{fig2}}
\end{figure}

These results mentioned above are shown in Figure~\ref{fig2}.
Data points in the figure are the required amounts of $\rm ^{44}Ti$ as
a function of its half-life, which are determined by the luminosity
evolution. Trapezoidal region and filled circles are its amounts
estimated by the analysis of the bolometric light curve
(\cite{mochizuki98}; \cite{mochizuki99a}).
It is noted that Mochizuki et al. (1999a) also calculated the required
amount for an effectively longer half-life ($\sim$ 100 yrs). They
pointed out the possibility that its half-life becomes 
longer if $\rm ^{44}Ti$ is highly ionized and the decay channel of the
orbital electron capture is blocked out in a supernova remnant
(\cite{mochizuki99a}), although the temperature has
to be $\sim 10000$ times higher than the observed one in SN 1987A
(\cite{haas90}) in order to attain such a highly ionization.
Open circle is that estimated by the analysis of the light curves for
the B and V bands (Kozma 1999). Down arrow is the upper limit on the
amount estimated by the flux of Fe[II] 25.99 $\rm \mu$m line
(\cite{lundqvist99}). The most reliable value for its half-life is
$\sim$ 60 yrs (Ahmad et al. 1998; G$\rm \ddot{o}$rres et al. 1998;
Norman et al. 1998). $\tau = 66.6$ yrs is
the upper limit of its half-life obtained by the experiments (Alburger
\& Harbottle 1990). Its lower limit obtained by the
experiments is $\tau = 39$ yrs (Meissner 1996).
Horizontal lines are results of the numerical calculations Ha95,
WW95, TNH96, and Na97. WW95 represents the model T18A in their paper,
since $\rm ^{44}Ti$ is synthesized most in this model among their
models in the mass range (18-21)$M_{\odot}$.

\placefigure{fig2}

\subsection{Results} \label{result1}
\indent

\begin{figure}
\epsscale{1.0}
\plottwo{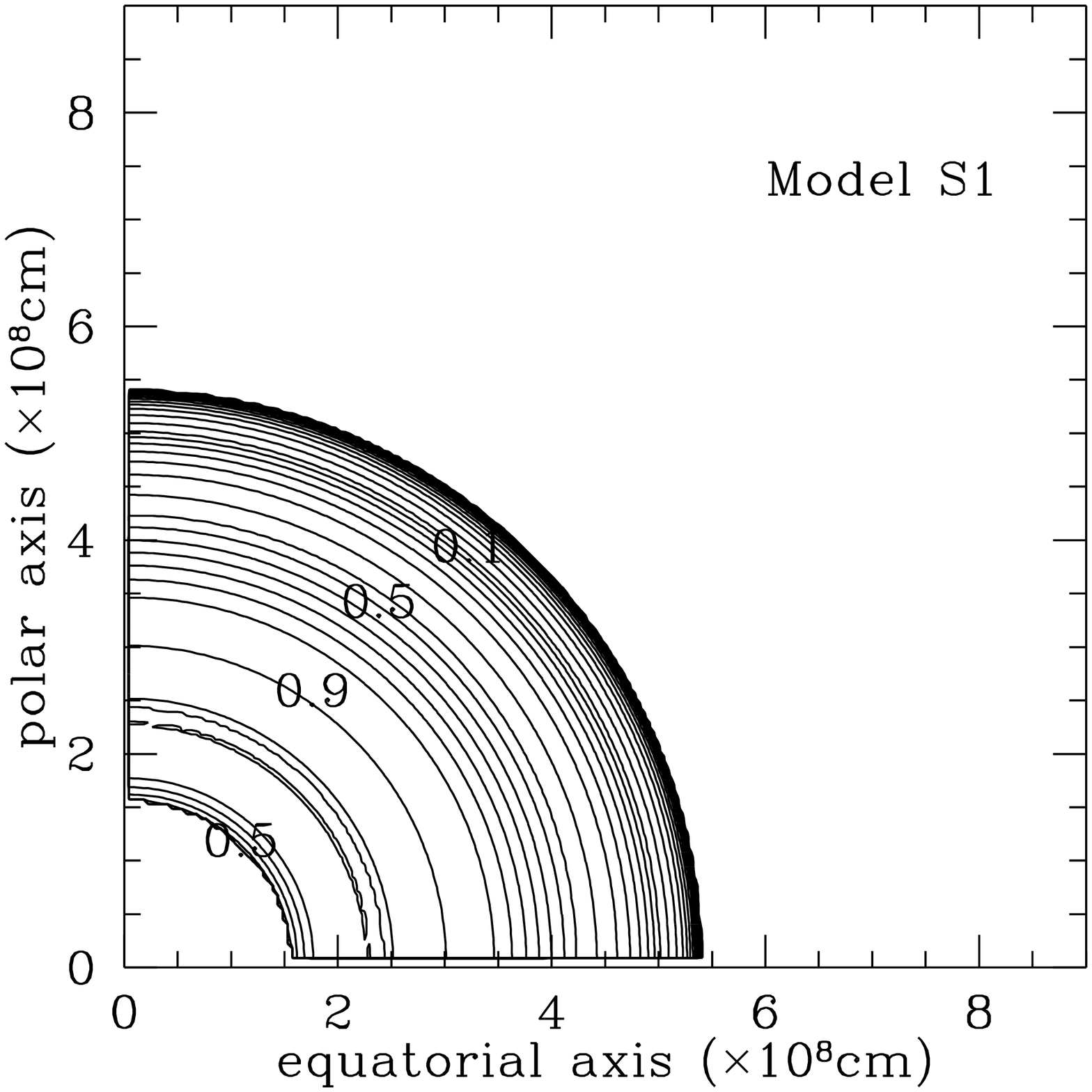}{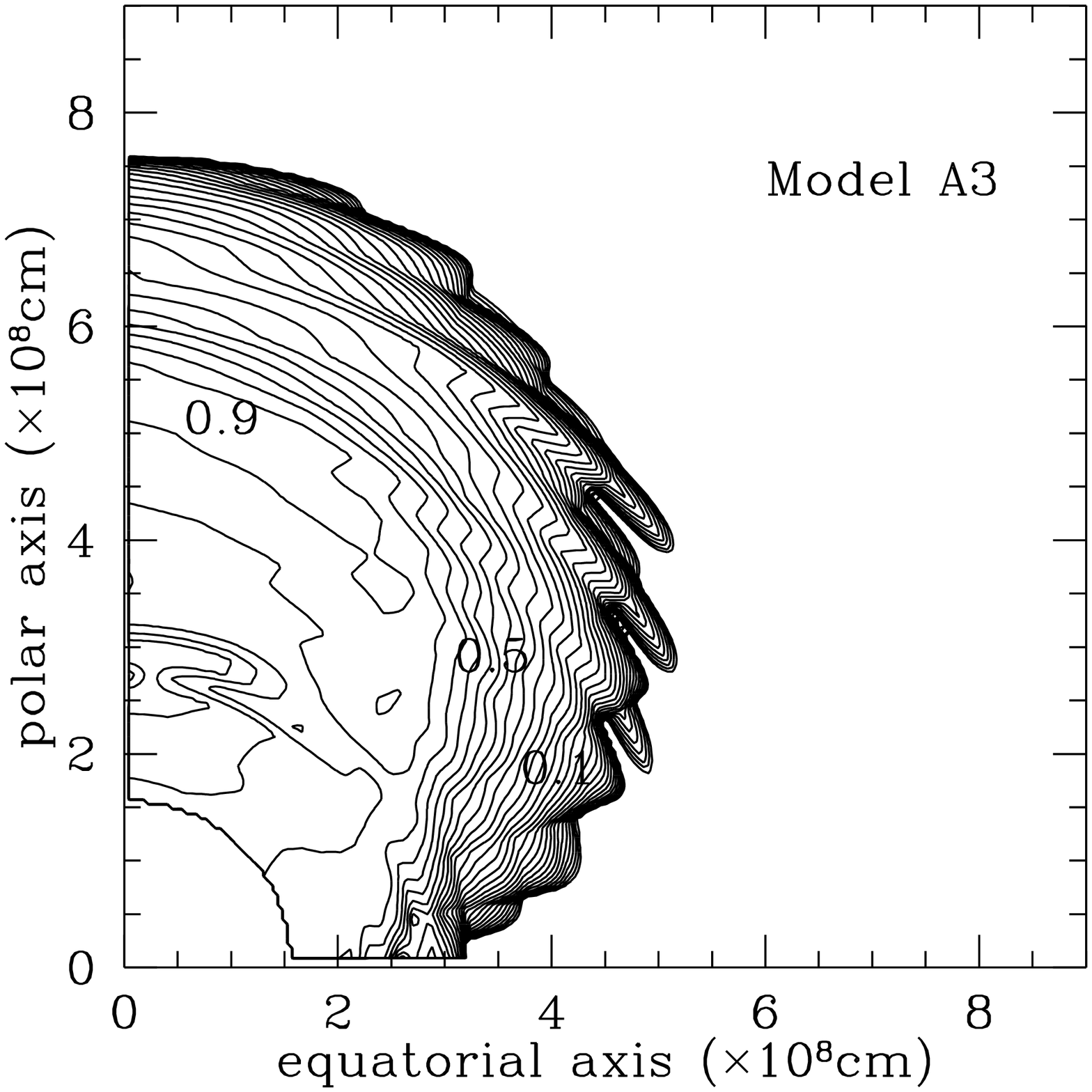}
\figcaption{
Contours of the mass fraction of $\rm ^{56}Ni$ after the explosive
nucleosynthesis. Contours are
drawn for the initial position of the matter in the progenitor. The
inner boundary at $r \sim 1.8 \times 10^{8}$ cm means the Si/Fe interface.
Left: the result for the model S1. The maximum value of the mass
fraction of $\rm ^{56}Ni$ is $9.3 \times 10^{-1}$. The regions where
the mass fraction of $\rm ^{56}Ni$ becomes 0.9, 0.5, and 0.1 are noted 
in the figure. Right: Same as left, but for the model S1. The maximum
value is $9.1 \times 10^{-1}$.
\label{fig3}}
\end{figure}

Results of the calculations of the explosive nucleosynthesis are shown 
in this subsection. First, the contours of the mass fraction of $\rm
^{56}Ni$ in a progenitor are shown in Figure~\ref{fig3}. Contours are
drawn for the initial position of the matter in the progenitor. The
inner boundary at $r \sim 1.8 \times 10^{8}$ cm means the Si/Fe interface.
Left is the result for the model S1 and right is the one for A3. As can
be seen from the figures, $\rm ^{56}Ni$ is synthesized at the inner
most region of the ejecta and is a good tool to determine the location 
of the mass cut (see subsection~\ref{masscut}). It is also noted that
$\rm ^{56}Ni$ is synthesized only in the polar region in the model A3.
This is because the shock wave in the polar region is strong and
temperature rise high sufficiently to synthesize much of iron-group
elements, while the shock wave around the equatorial plane is too weak to
synthesize them. This fact has a great influence on the line profiles
of Fe$\rm [II]$ in the ejecta. This is discussed in
subsection~\ref{result2} and~\ref{discussion2}.

\placefigure{fig3}

\begin{figure}
\epsscale{1.0}
\plottwo{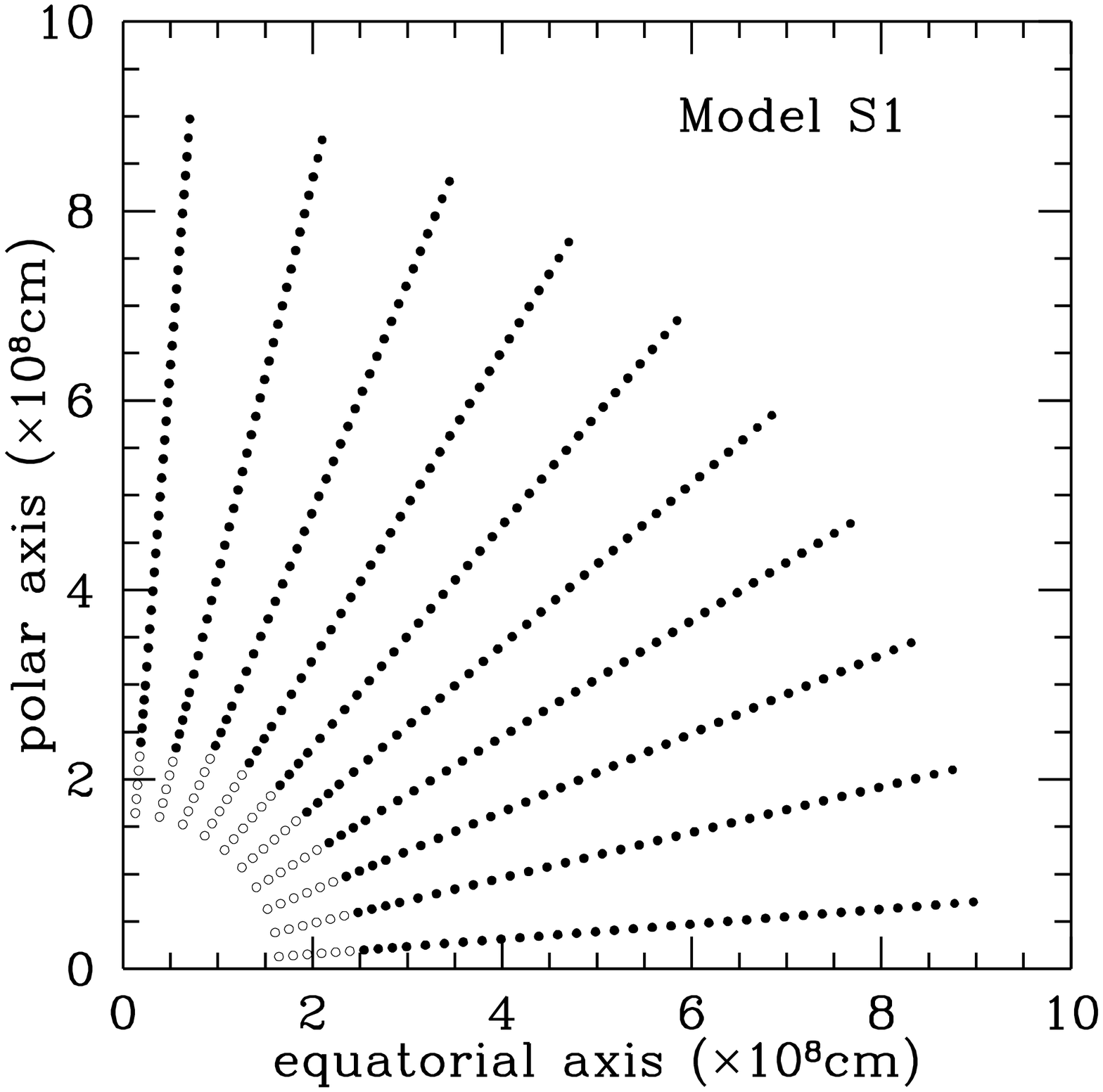}{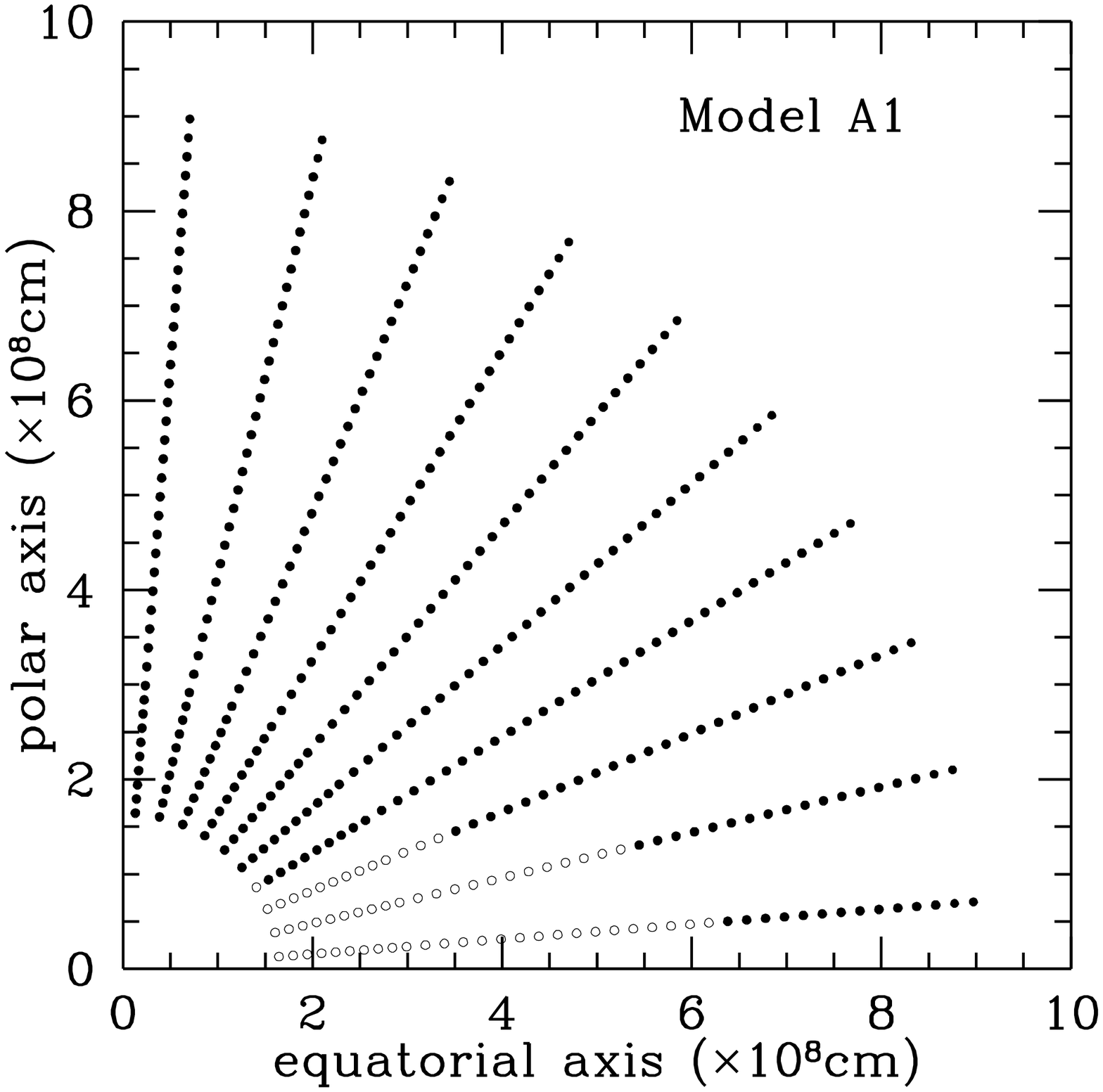}
\figcaption{
Forms of asymmetric mass cut A7.
Filled and open circles represent the test particles that will be
ejected and falling back, respectively. A mass cut is defined as an
interface of filled/open circles. Left: for the model S1; right: for
the model A1.
\label{fig4}}
\end{figure}

\begin{figure}
\epsscale{1.0}
\plottwo{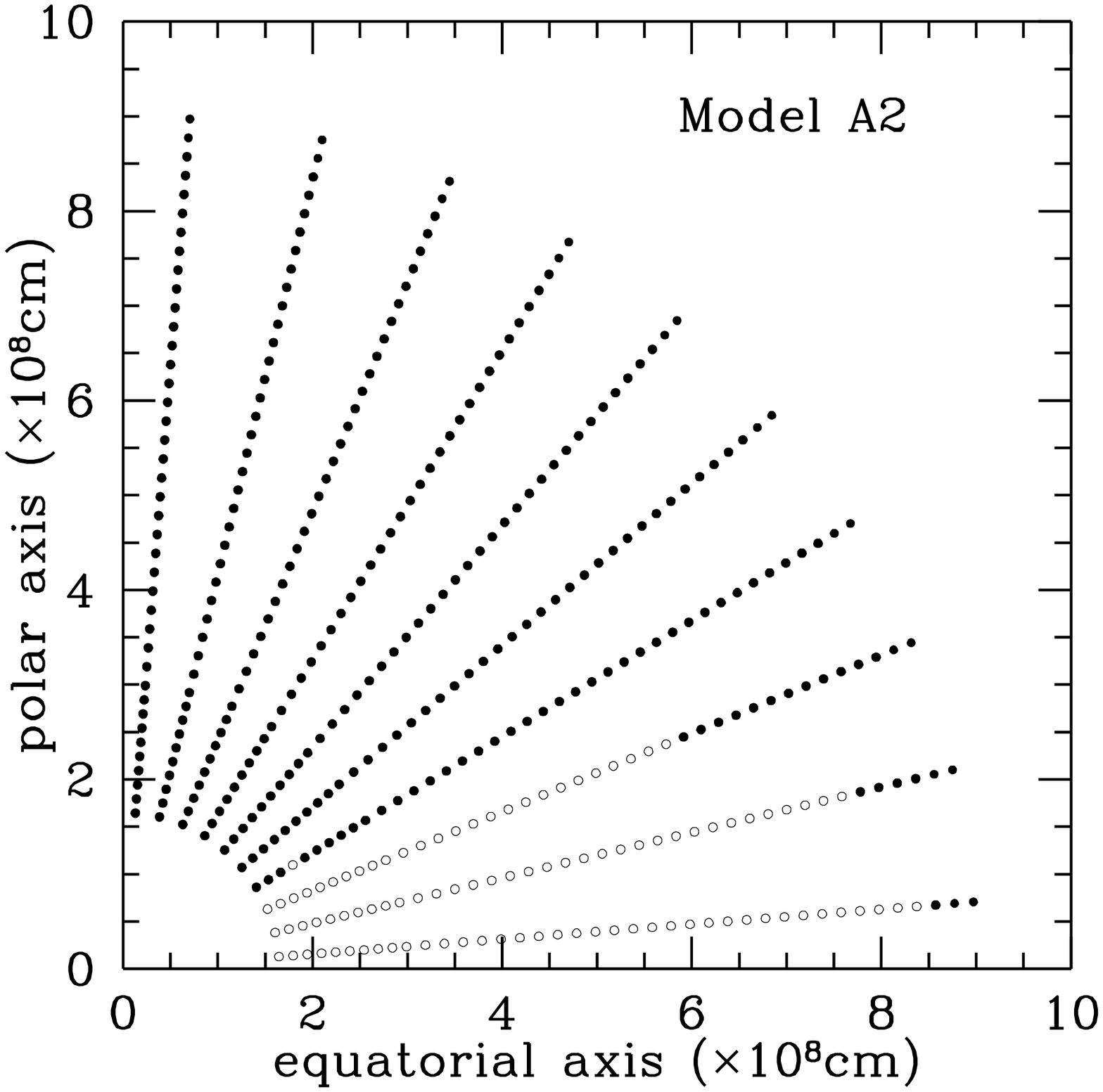}{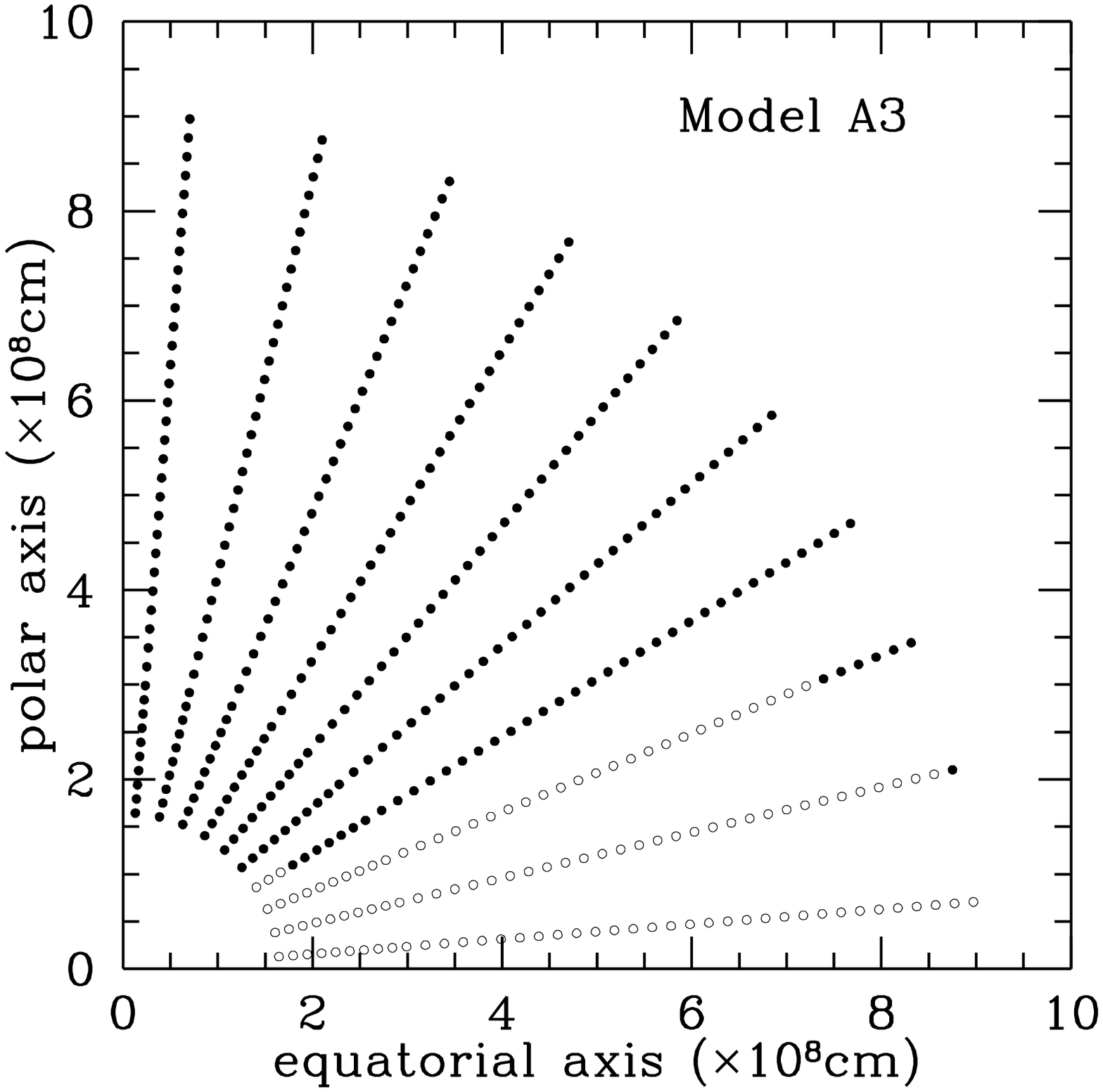}
\figcaption{
Same as figure 4, but for the models A2 (Left), and A3 (Right).
\label{fig5}}
\end{figure}

Forms of the mass cut A7 are presented in Figures~\ref{fig4} and~\ref{fig5}.
Filled and open circles represent the test particles that will be
ejected and falling back, respectively. A mass cut is defined as an
interface of filled/open circles. These are plotted for their initial
positions. We can find the tendency that the 
matter in the polar region is ejected more than that around the equatorial 
plane. This can be easily understood because the strong shock wave in
the polar region gives more energy to the matter. These figures prove
that the guiding principle to determine the location of the mass cut
presented in subsection~\ref{masscut} is not so bad. This problem is also
discussed in subsection~\ref{result2} and~\ref{discussion2}.

\placefigure{fig4}
\placefigure{fig5}

\begin{figure}
\epsscale{1.0}
\plottwo{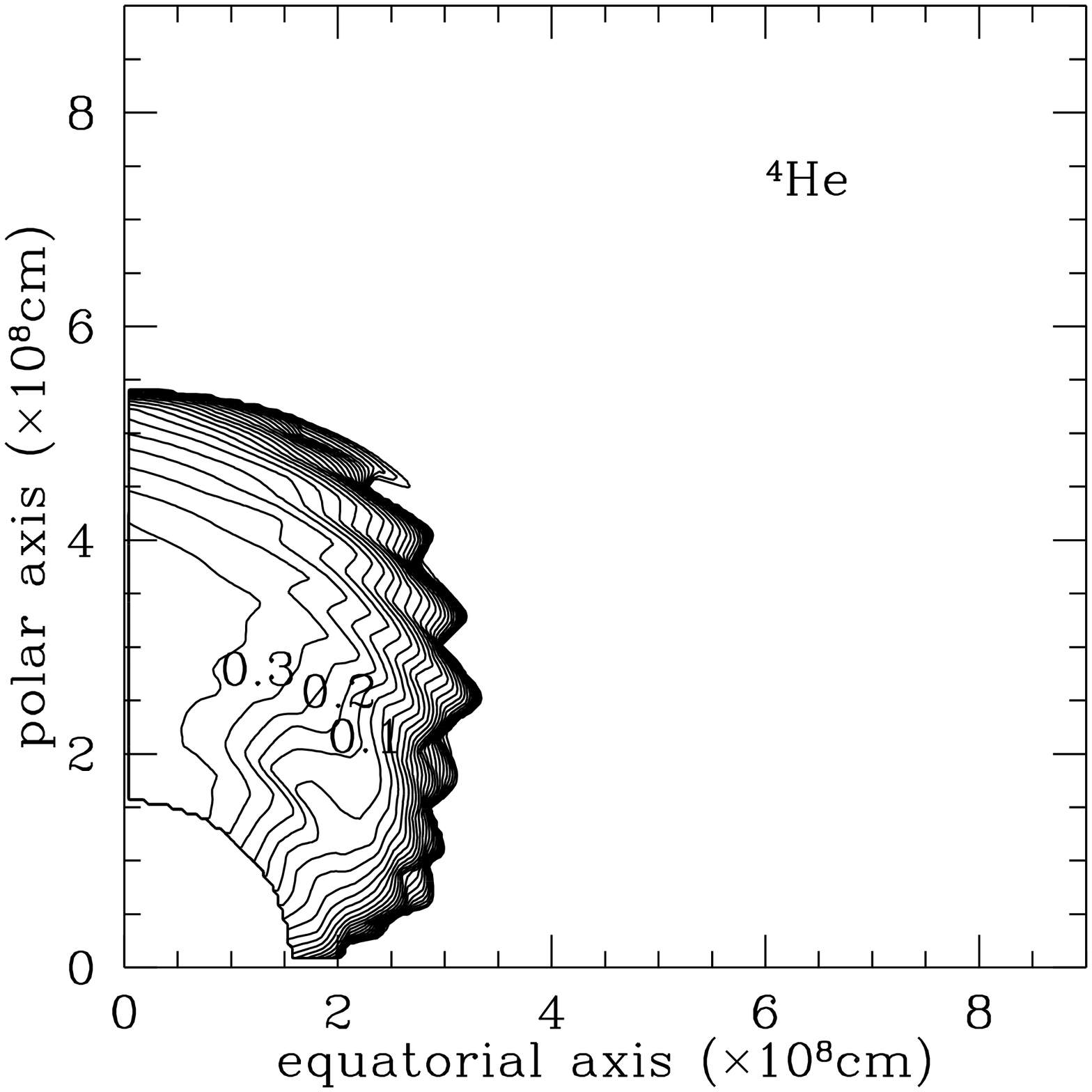}{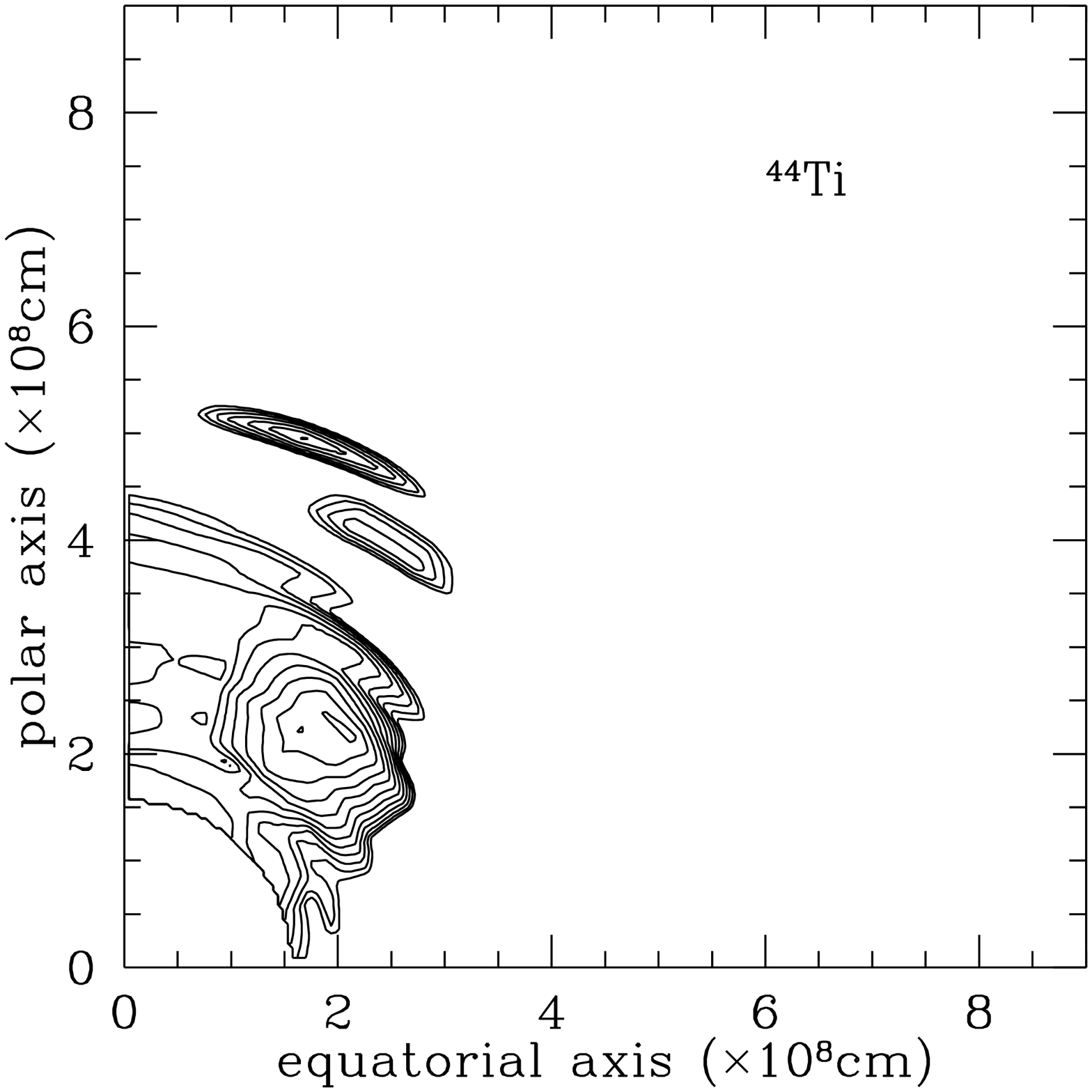}
\figcaption{
Contours of the mass fraction of $\rm ^{4} He$ and $^{44}\rm Ti$ after 
the explosive nucleosynthesis in the model A3. Left: Contours of the
mass fraction of $\rm ^{4} He$. Right: Contours of the mass fraction of 
$^{44}\rm Ti$. Maximum values of the mass fraction of $\rm ^{4}He$ and 
$\rm ^{44}Ti$ are $4.0 \times 10^{-1}$ and $1.3 \times 10^{-2}$.
The regions where the mass fraction of $\rm ^{4}He$ becomes 0.3,
0.2, and 0.1 are noted in the left figure.
\label{fig6}}
\end{figure}

Contours of the mass fraction of $\rm ^{4} He$ and $^{44}\rm Ti$ after 
the explosive nucleosynthesis in the model A3 are drawn in
Figure~\ref{fig6}. Much of $\rm ^{4}He$ is synthesized near the polar
axis for the jet-like explosion models since a high photon to baryon ratio is
achieved there. That is, photons with high energy that decompose heavy 
elements into light nuclei are filled in the polar region, which
causes the alpha-rich freezeout (\cite{thielemann96}; \cite{nagataki97}).
Since $\rm ^{44}Ti$ is produced through this process (\cite{the98}),
we can find the correlation between the contours of $\rm ^{44}Ti$ and
$\rm ^{4}He$.

\placefigure{fig6}

\begin{figure}
\epsscale{1.0}
\plotone{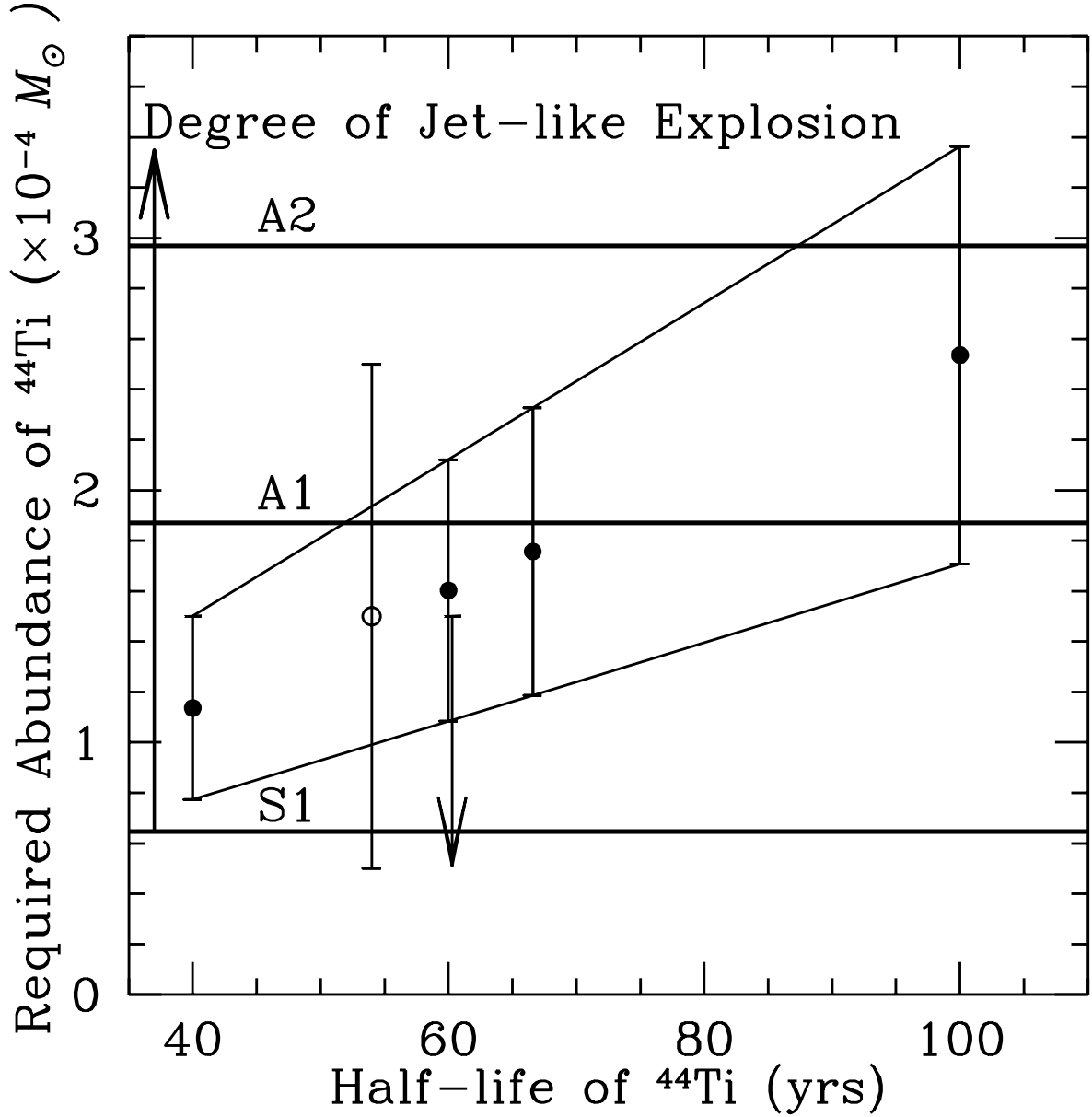}
\figcaption{
Calculated masses of $\rm ^{44}Ti$ by the S1, A1, and A2 models.
Asymmetric mass cut A7 is adopted. The synthesized mass of $\rm ^{44}Ti$
becomes larger along with the degree of the jet-like explosion.
Required amounts of $\rm ^{44}Ti$ are
also shown as a function of its half-life (Mochizuki \& Kumagai
1998; Mochizuki et al. 1999a; Kozma 1999; Lundqvist et al. 1999).
The most reliable value
for its half-life is $\sim$ 60 yrs (Ahmad et al. 1998; G$\rm
\ddot{o}$rres et al. 1998; Norman et al. 1998).
\label{fig7}}
\end{figure}

Calculated masses of $\rm ^{44}Ti$ by the S1, A1, and A2 models
are shown in Figure~\ref{fig7}.
We can see the tendency that the synthesized mass of $\rm ^{44}Ti$
becomes larger along with the degree of the jet-like explosion. The
reason for it is mentioned above. This tendency does not depend on the 
form of the mass cut (see Table~\ref{tab4}). It is noted that
the model A1 is a good one to explain the amount of $\rm ^{44}Ti$ in SN
1987A. Additionally, even if the effective half-life of $\rm ^{44}Ti$
is longer than that in a laboratory (\cite{mochizuki99a}), the
jet-like explosion model can meet the requirement.

\placefigure{fig7}

As for the amounts of $\rm ^{57}Ni$ and $\rm ^{58}Ni$, calculated
amounts of these nuclei are summarized in Table~\ref{tab4}.
The calculated ratios of $\rm ^{57}Ni / ^{56}Ni$ are in agreement with 
the observed ratio, while the calculated ratios of $\rm ^{58}Ni /
^{56}Ni$ seem to be a little higher than the observed value. Since
these ratios does not sensitive to the form of the initial shock wave
and of the mass cut, these ratios may depend mainly on the progenitor
model (\cite{nagataki98c}). 

\placetable{tab4}

\subsection{Discussion on the Explosive Nucleosynthesis} \label{discussion1}
\indent

A lot of discussions have already been presented in the previous
subsections. Additional comments are presented in this subsection.
We have shown that $\rm ^{44}Ti$ can be synthesized in a sufficiently 
amount to explain the observed luminosity in SN 1987A around 3600 days
after the explosion, when a jet-like explosion is assumed. In
particular, it is shown that the model A1 is a good one to
explain the observed luminosity. There are other candidates that may
explain the luminosity. One is a pulsar activity (\cite{kumagai93})
and the other is the freezeout effect (\cite{fransson93}). The
freezeout effect means that the time scale of the recombination of the
plasma can not be ignored and the assumption of the instantaneous energy
input due to the recombination breaks down. We think that the pulsar
activity is a promising candidate, although a pulsar has not been
found in SN 1987A yet and its activity has not been measured. The
freezeout effect is also interesting one, although there are
uncertainties and difficulties in making a model for the structure of
the ejecta (\cite{fransson93}). Moreover, it is reported that this freeze-out
effect becomes less important as $\rm ^{44}Ti$ starts to dominate
(\cite{lundqvist99}).
The contribution and amount of $\rm ^{44}Ti$ will be determined
clearly when the ejecta becomes optically thin and $\gamma$-ray line
is detected directly, although it will take $\sim 100$ yrs.

It is also noted that $\gamma$-ray line from the decay of $\rm
^{44}Ti$ has been detected in Cassiopeia A, which is famous for the
asymmetric form of the remnant. The inferred abundance ratio of $\rm
^{44}Ti$ relative to $\rm ^{56}Ni$ is quite high, as a jet-like
explosion model predicts (\cite{nagataki98d}).

Such a high entropy condition caused by a jet-like explosion will be
also good for the synthesis of the rapid-process nuclei
in the iron core (\cite{woosley94}; \cite{mclaughlin96}).
We note that the nucleosynthesis in the iron core is not calculated in 
this study, although such a calculation is now underway. We
will report the results in the near future.
Additionally, a hypernova model (\cite{iwamoto98}; \cite{woosley99})
will be also able to realize such a high entropy condition.

Finally, we emphasize again that we have found the favored degree of a
jet-like explosion to explain the tail of the light curve. In the next 
section, we show this degree is also good for explaining the line profiles
of Fe$\rm [II]$ in SN 1987A.

\section{MATTER MIXING IN SN 1987A}\label{mixing}
\indent

\subsection{What are the Problems?} \label{problem2}
\indent

SN 1987A also provided us the apparent evidence of large-scale mixing
in the ejecta for the first time.
For example, the early detection of X-rays
(\cite{dotani87}; \cite{sunyaev87}; \cite{wilson88}) and $\gamma$-rays
(\cite{matz88}) from the radioactive nuclei $\rm ^{56}Co$ reveals that 
a small fraction of the matter at the bottom of the ejecta was mixed
up to the outer layer. It should be also noted that these early
detections verified the prediction derived by Clayton (1974), which
had been discussed 13 years before SN 1987A exploded.
The form of the X-ray light curve during (100-700) days after the
explosion is also thought to be the indirect evidence of mixing and
clumping of the heavy elements (\cite{itoh87}; \cite{kumagai88}).
Moreover, the observed infrared line profiles of heavy
elements, such as $\rm Fe [II]$, $\rm Ni [II]$, $\rm
Ar [II]$, and $\rm Co [II]$, show that a part of them is
mixed up to the fast moving (3000-4000 km/s) outer layers
(\cite{erickson88}).

At present, the growth of a fluctuation in the progenitor due to the
Rayleigh-Taylor (R-T) instability is thought to be the most promising
scenario for the matter mixing. Although this idea was not a new one
(e.g., \cite{falk73}; \cite{chevalier76}; \cite{bandiera84}), it was
necessary to perform multi-dimensional hydrodynamic calculations with
a realistic stellar model including the fluctuations in order to
investigate their growth quantitatively.
With the improvement of the supercomputers, many people have done such
calculations. They performed mainly 2-dimensional calculations of the
first few hours of the explosion. As a result, they showed that the
fluctuations indeed grow due to the R-T instability (\cite{arnett89};
\cite{hachisu90}; \cite{muller90}; \cite{fryxell91}; \cite{muller91}).

However, there are some points which are still open to arguments.
One problem is how and where the initial fluctuations, the seed of the
matter mixing, are made in a progenitor. Also, their amplitude has not been
known yet. Up to the present, two candidates have been proposed for
that seed.
One idea is that these fluctuations are generated by the convection
during the stellar evolution. It is reported that the
density fluctuation at the inner and outer boundaries of the
convective O-- rich layer, $\delta \rho / \rho$, becomes $\sim 5 \%$
(up to $8 \%$) at the beginning of the core collapse (\cite{bazan94};
\cite{bazan97}).
The other is that the initial fluctuations are amplified when the
shock wave is formed in the iron core. It is reported that the
amplitude of the fluctuations, $\delta R_s / R_s$ ($R_s$ is the shock
radius), becomes $\sim 30 \%$ in the iron core (\cite{burrows95}),
even if the initial fluctuations are set to be small ($\sim 1-2 \%$).

Another problem is the reproduction of the observed line profiles of
heavy elements.
As mentioned above, the line profiles of Fe$\rm  [II]$ have shown
that a small fraction of Fe is traveling at (3000-4000) km/s. 
On the other hand, the velocities of Fe are of order 2000 km/s at most 
in the numerical simulations, even if the acceleration by the energy
release of the radioactive nuclei is taken into account (\cite{herant91};
\cite{herant92}). Although they insist that pre-mixing of $\rm
^{56}Ni$ will be necessary for the reproduction, this problem
seems to be unresolved. It is also noted that $2 \times 10^{51}$ ergs
are required for an explosion energy in order to reproduce the fast moving
component in their model.
As for the grobal form of the line profiles, the observed line
profiles are asymmetric with a steep edge on the redshifted side and a
more gradual decline on the blueshifted side. Maximum flux density is
also redshifted. This feature has not been reproduced yet by
the numerical simulations.

In this section, the effects of jet-like explosion on the matter mixing
are investigated. It is noted that Yamada $\&$ Sato (1991) did
2-dimensional hydrodynamic calculations assuming a jet-like
explosion. They found that heavy elements could be highly accelerated
in a jet-like explosion and get velocities of order of 4000km/s when
the amplitude of the initial fluctuations is set to be $30\%$.
However, there are some points to be improved in their calculations.
For example, only one model is assumed for the calculation of the jet-like
explosion. The initial velocity behind the shock wave is assumed to be
proportional to $r \times \cos^{2} \theta$, where $\theta$ is the zenith
angle. They did not calculate the line profiles of heavy elements,
which should be compared with the observations. They also assumed that
the chemical composition of the ejecta and the form of the mass cut are
spherically symmetric. In this study, we improve their calculations
and compare the results with the observations. Our main aim is, of
course, to find the proper degree of the jet-like explosion in SN 1987A.

\subsection{Line profiles of Fe$\rm [II]$}\label{required2}
\indent

\begin{figure}
\epsscale{1.0}
\plottwo{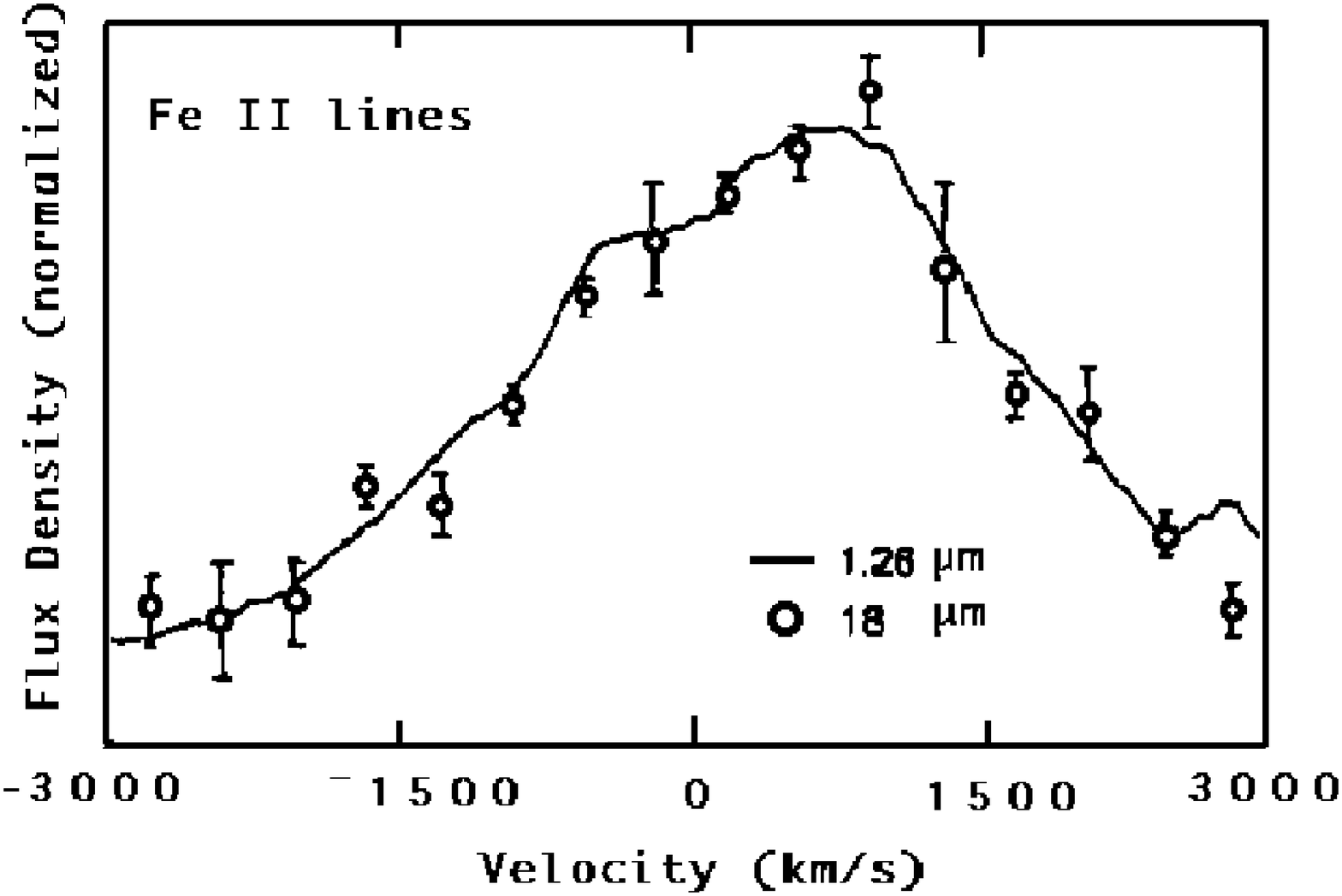}{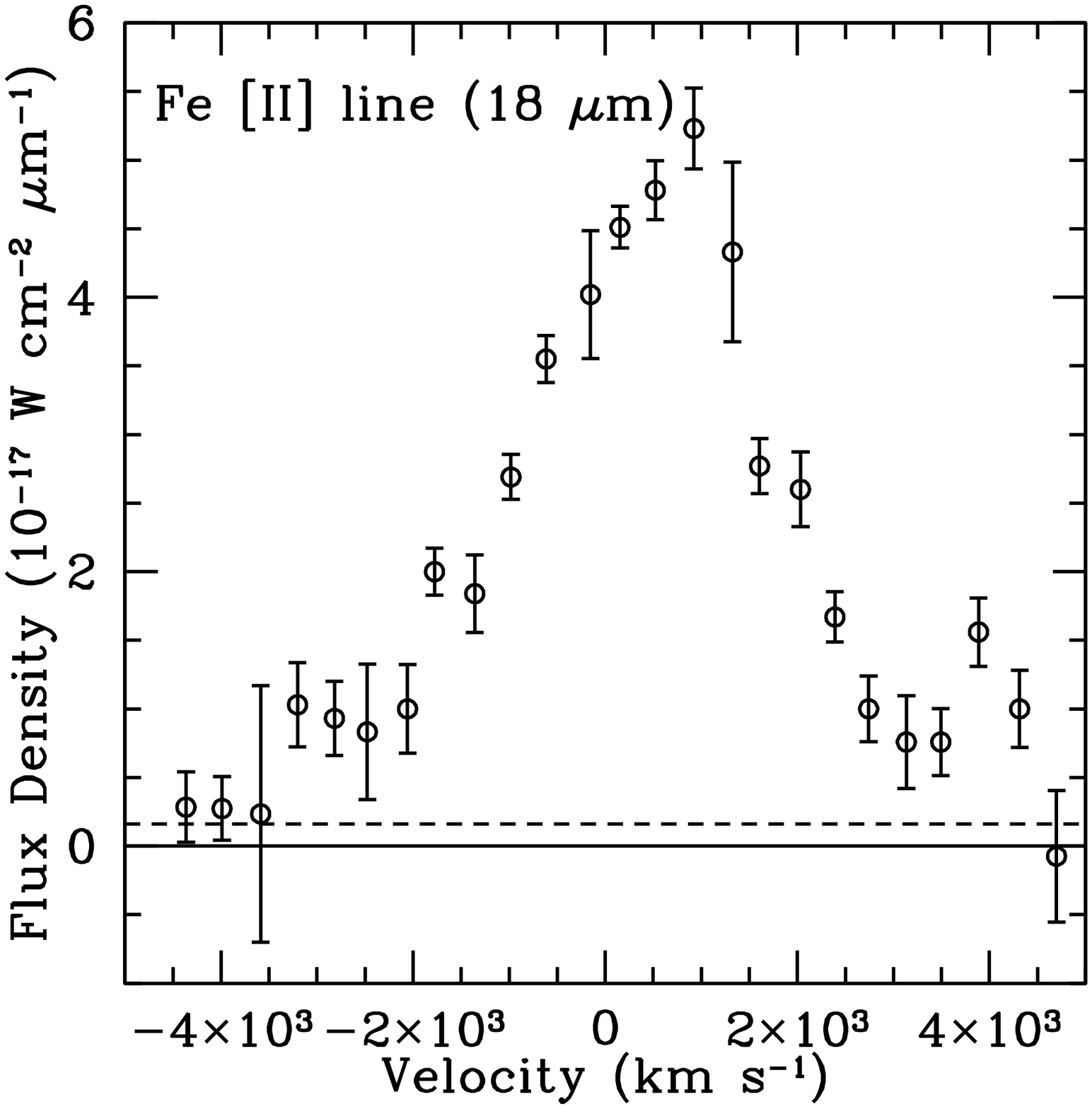}
\figcaption{
Observed line profiles of Fe$\rm [II]$. Left: comparison of the
central portion of the 18$\rm \mu$m profile at 409 days after the
explosion (Haas et al. 1990) with the 1.26 $\rm \mu$m profile at 377
days (Spyromilio et al. 1990). Positive velocity corresponds to a
red-shift one. Right: Line profile of the 18$\rm \mu$m profile covering 
$\pm 4500$ km/s. The 1 $\sigma$ error bars are included. Adopted
continuum level is indicated by dashed line.
\label{fig8}}
\end{figure}

Observed line profiles of Fe$\rm [II]$ are shown in
Figure~\ref{fig8}. A comparison of the central portion of the 18$\rm
\mu$m profile at 409 days after the explosion (\cite{haas90}) with the
1.26 $\rm \mu$m profile at 377 days (\cite{spyromilio90}) is shown in
the left panel. Positive velocity corresponds to a red-shift one.
Although the data for the 1.26 $\rm \mu$m spectrum covers $\pm$
3000 km/s, those for the 18$\rm \mu$m one covers $\pm$ 4500 km/s,
which is shown in the right panel. Adopted continuum level is indicated 
by dashed line.

\placefigure{fig8}

Several comments on the figures will be needed. First, the iron in the 
ejecta is mainly produced through the decays of $\rm ^{56}Ni
\rightarrow ^{56}Co \rightarrow ^{56}Fe$. Also, the bulk of the
observed iron
is Fe$\rm [II]$, as deduced from infrared transitions of Ni$\rm [I]$,
Ni$\rm [II]$, and Ni$\rm [III]$ (\cite{haas90}). So we compare the velocity
distribution of $\rm ^{56}Ni$ in the numerical calculations with the
observed line profiles of Fe$\rm [II]$.
Second, we can find clearly that a small fraction of Fe is
traveling at (3000-4000) km/s, which can not be reproduced by the
spherical explosion models.
Third, it is an important fact that the line
profiles of different wave lengths share the same form of the
spectrum, which shows the
optical depth of the emission region is quite low. This is because the
dependence of the optical depth on the wave length is quite large when
the density of the system is high and the optical depth is
not zero (\cite{spyromilio90}).
Fourth, there is an apparent asymmetry of the form of the line
profiles. In particular, the flux from the red-shift side is higher
than that from the blue-shift side, which is contrary to the intuition.
Spyromilio et al. (1990) attributed this line asymmetry primarily to
emission from regions where density and/or electron temperature depart 
from spherical symmetry. Furthermore, Haas et al. (1990) concluded
that the asymmetry results from the density inhomogeneities, rather
than the temperature ones, because of the extreme temperature
sensitivity of the higher excitation 1.26 $\mu$m line.
Finally, Haas et al. (1990) used their
radiation transfer models and concluded that the Fe$\rm [II]$ lines
are partially optically thick and the missing iron is
probably hidden in regions of enhanced optical depth or 'clumps',
rather than in a smooth density distribution with higher optical
depths at low velocities.

In this study, we compare the velocity distribution of $\rm ^{56}Ni$
in the numerical calculations with the observed line profiles
assuming they are optically thin (\cite{herant92}).
This can be justified as follows. As
for the reproduction of the fast moving component, it is important to
show the fact that it can be produced in the jet-like explosion
models. We emphasize again that it can not be reproduced by the
spherical explosion models. So this fact is important, irrespective of
the optical depth. As for the form of the line
profiles, we use it to make a constraint on the degree of the jet-like 
explosion in this study. It is true that the form of the observed line
profiles may change and the constraint on the degree may be also changed 
when the missing iron in the clumps becomes to appear. However, as
shown in subsection~\ref{result2}, the calculated velocity
distributions with large $\alpha$ are far from similar to the form
of the observed profiles. That is why we think our conclusion in this
study is not so sensitive to the presence of the missing iron.
Such insensitivity will be also supported by the conclusion of Haas et 
al. (1990) that the missing iron is not hidden as a sphere at the
center of the system but as clumps. This is because the form of the line
profiles will not change so much when the missing iron in the clumps
becomes to appear.

\subsection{Results} \label{result2}
\indent

\begin{figure}
\epsscale{1.0}
\plotone{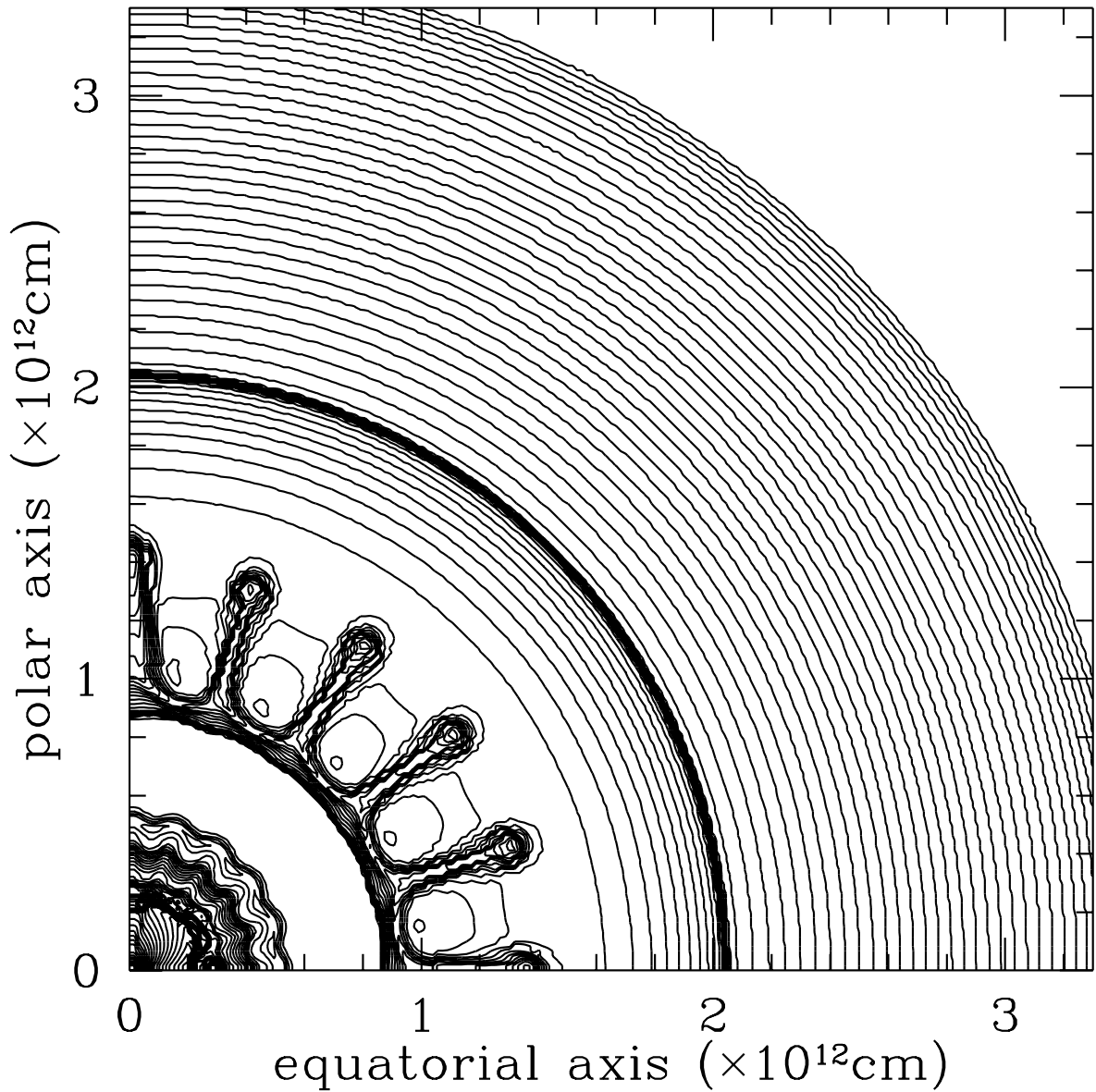}
\figcaption{
Density contours for the model S1b at $t$ = 5000 s after the explosion.
The radius of the surface of the progenitor is
3.3 $\times 10^{12}$ cm. The shock wave can be seen at the radius
$\sim 2 \times 10^{12}$ cm. Hydrogen envelope is lying during (1-3.3)
$\times 10^{12}$ cm. Heavy elements, such as Ni, Fe, Si, O, and C, are
packed in the layer lying $\sim 1 \times 10^{12}$ cm. The growth of
the fluctuations due to the R-T instability can be seen behind the shock wave.
\label{fig9}}
\end{figure}

Results of the calculations of the matter mixing are shown in this
subsection. Density contours for the model S1b at $t$ = 5000 s after
the explosion are shown 
in Figure~\ref{fig9}, which tells us how the matter mixing is going on.
The radius of the surface of the progenitor is
3.3 $\times 10^{12}$ cm. The shock wave can be seen at the radius
$\sim 2 \times 10^{12}$ cm. Hydrogen envelope is lying during (1-3.3)
$\times 10^{12}$ cm. On the other hand, heavy elements, such as Ni,
Fe, Si, O, and C, are packed in the layer lying $\sim 1 \times 10^{12}$ 
cm. Additionally, a weak flow of the matter is introduced at the inner
most boundary (= $10^{8}$ cm) in order to maintain the stability of
the hydrodynamic calculations. The region behind the layer that
contains heavy elements is
filled with this artificial matter. It is noted that the total energy
that the artificial matter has is much smaller than the explosion
energy ($\sim 1 \times 10^{51}$ ergs) and can be neglected.
The growth of the fluctuations due to the R-T instability can be seen
behind the shock wave. The amplitude of the fluctuations is, indeed, larger
than the initial one (= $5 \%$). We can find that the heavy elements
are conveyed to the outer layer by the instability.
 
\placefigure{fig9}

\begin{figure}
\epsscale{1.0}
\plotone{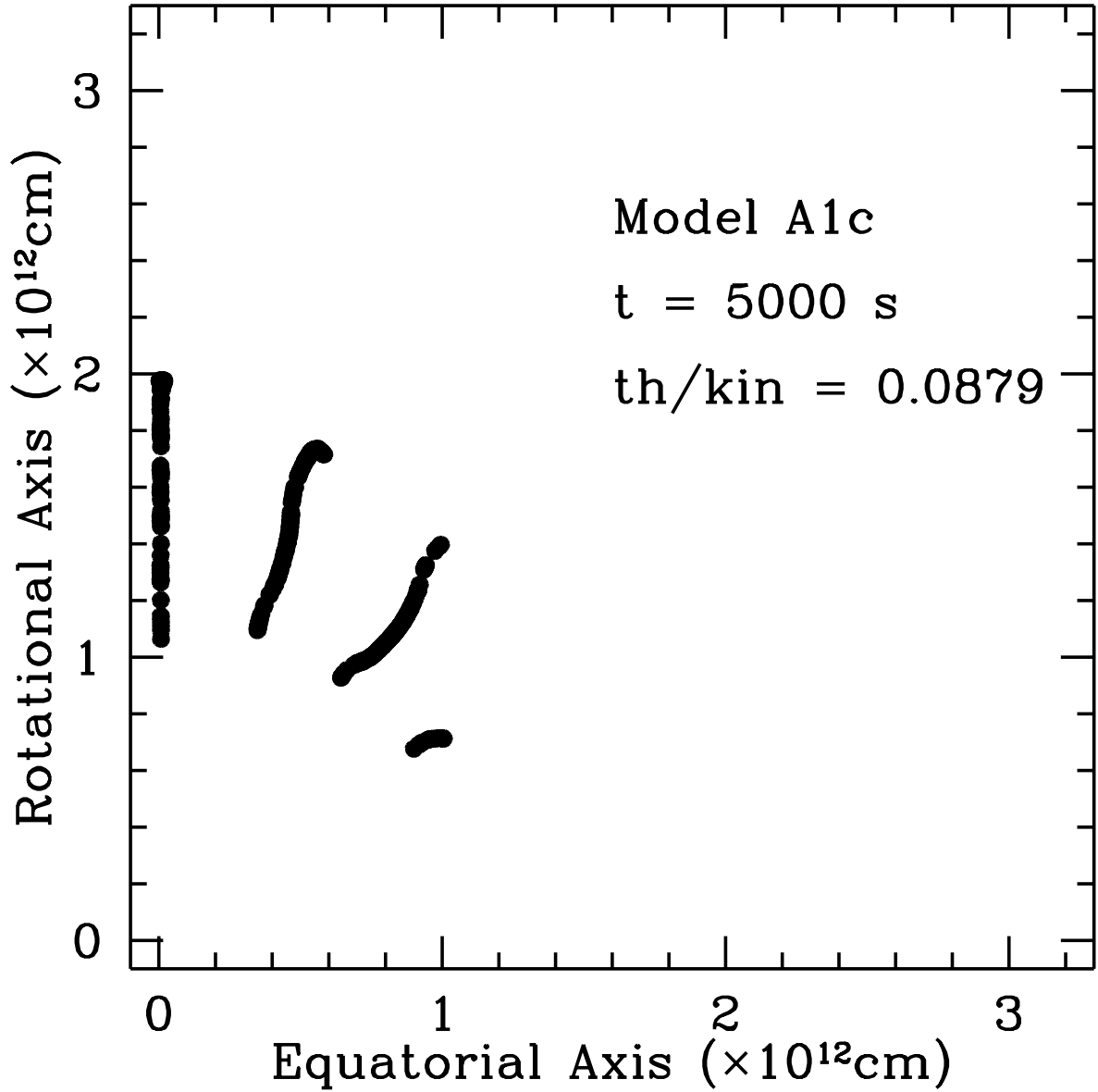}
\figcaption{
Positions of the test particles that meet the following
conditions for the model A1c at $t$ = 5000 s after the
explosion. The conditions are: (i) the mass fraction of $\rm ^{56}Ni$
is larger than 0.1 and (ii) velocity is higher than 2000 km/s. 
\label{fig10}}
\end{figure}

\begin{figure}
\epsscale{1.0}
\plottwo{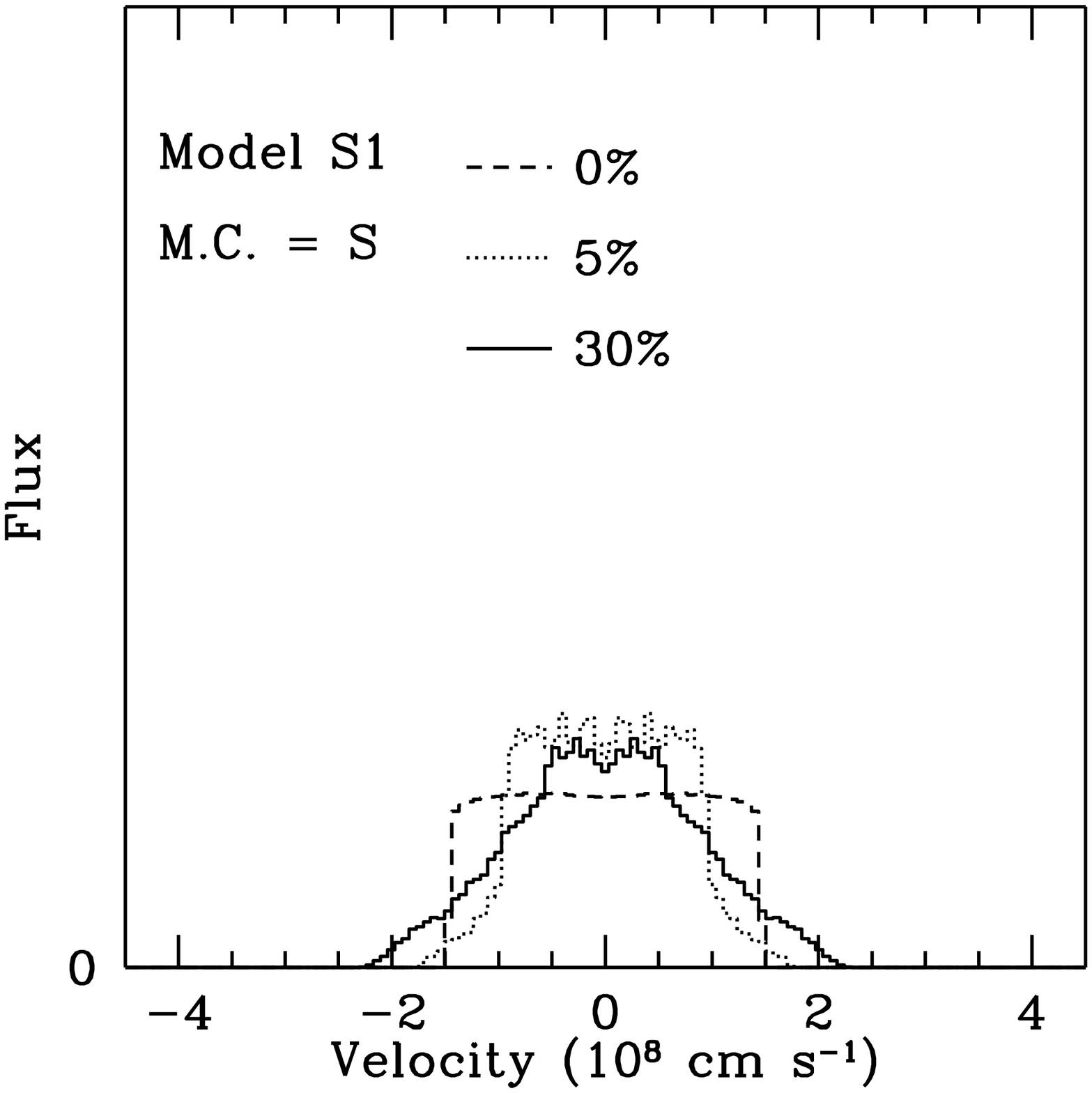}{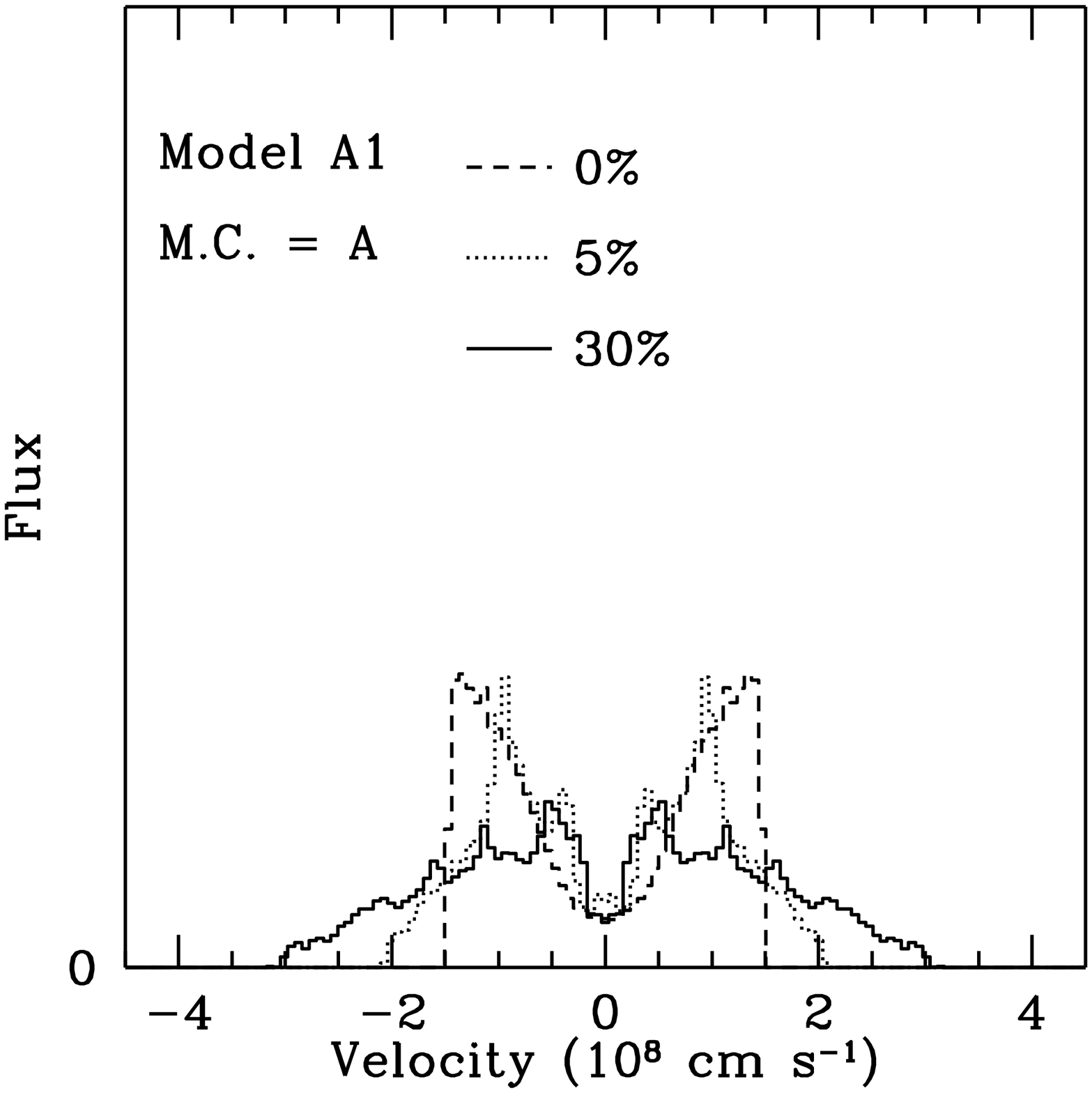}
\figcaption{
Velocity distributions of $\rm ^{56}Ni$ seen from $\theta =
44^{\circ}$ at $t$ = 5000 s after the explosion. Left: model S1;
right: model A1. Asymmetric mass cut A7 is adopted. The initial
amplitudes of the velocity fluctuations are $0 \%$ (short-dashed
curve), $5 \%$ (dotted curve), and $30 \%$ (solid curve), respectively.
\label{fig11}}
\end{figure}

In Figure~\ref{fig10}, we show the final positions ($t$ = 5000 s) of
the test particles that meet the following conditions for the model
A1c, which is concluded to be the best one in this study.
The conditions are: (i) the mass
fraction of $\rm ^{56}Ni$ is larger than 0.1 and (ii) velocity is
higher than 2000 km/s. As is clear from the figure, fast moving
component of $\rm ^{56}Ni$ is concentrated in the polar region, which
reflects the feature of the jet-like explosion. In particular, the
matter at the tip of the mushroom-like 'fingers' in the polar region
travels with a velocity higher than 3000 km/s, which is required from
the observations. We have to give an additional comment. Numerical
calculations in this study are performed until $t$ = 5000 s after the
explosion, which are compared with the observations at $t \sim 400$
days after the explosion. This is because such a wide region can not
be covered by an Eulerian coordinate used in this study. However,
there are some reasons that justify our treatment. One is that the
total internal energy relative to the total
kinetic energy in the heavy element layer is quite small (= 0.0879),
which means that the matter
in this layer expands almost freely already at $t$ = 5000 s. Second,
calculated velocity distributions of $\rm ^{56}Ni$ assuming a
spherical explosion in this study (see Figure~\ref{fig11}) are similar
to those in Herant et al. (1992), which are calculated
until $t$ = 90 days using a smooth particle hydrodynamics
code and assuming a spherical explosion (see their figure 9).

\placefigure{fig10}

\begin{figure}
\epsscale{1.0}
\plottwo{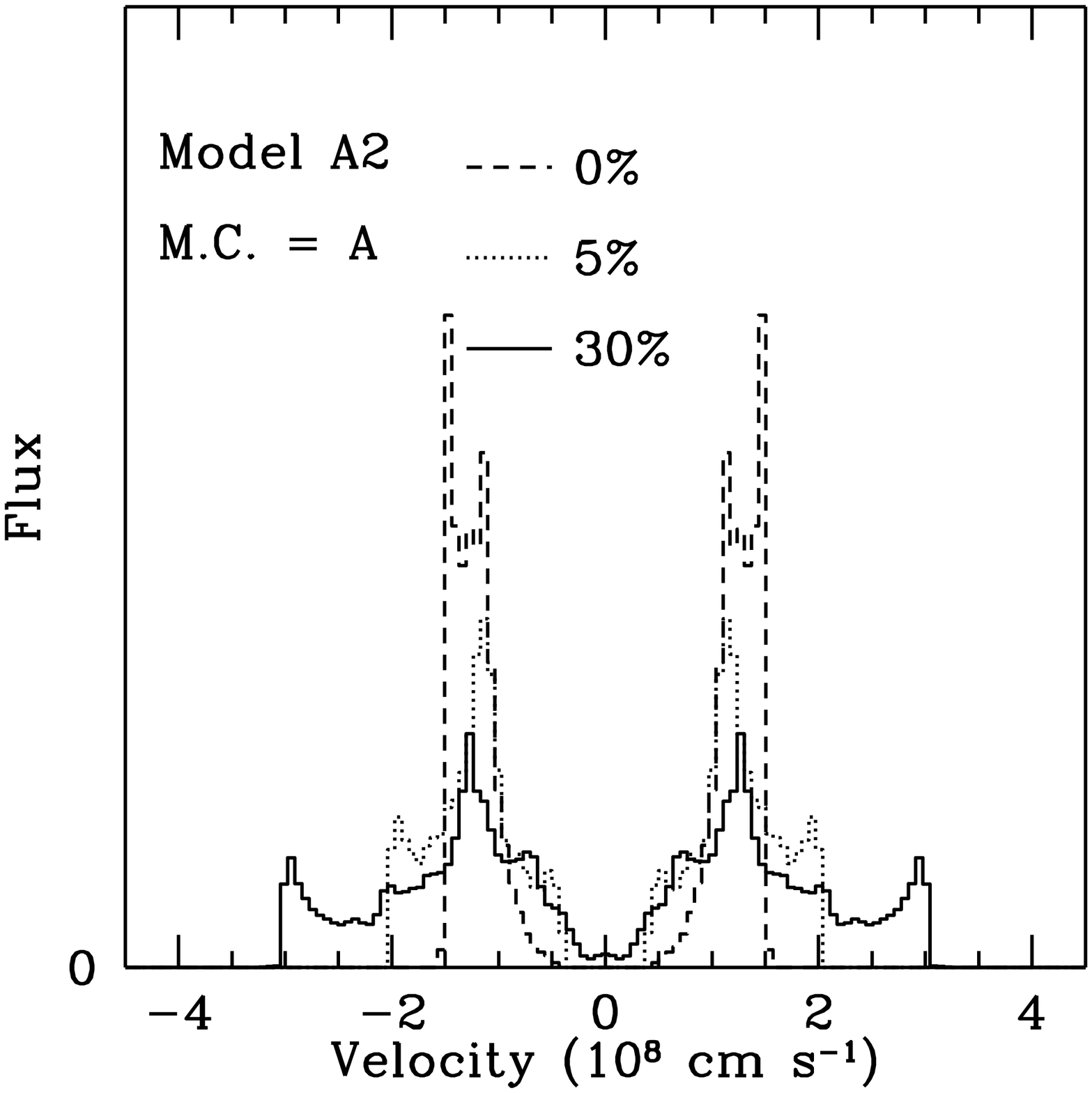}{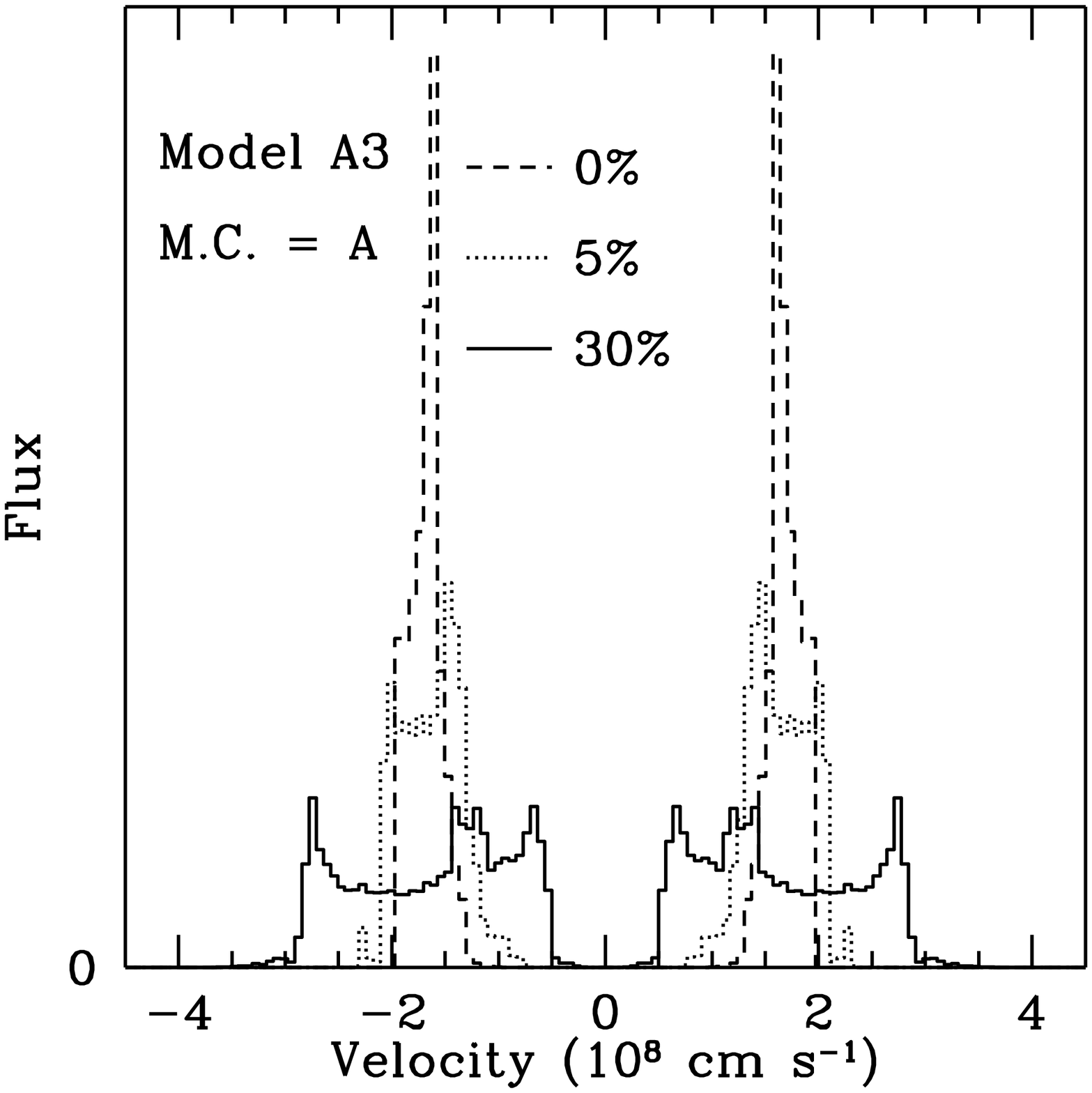}
\figcaption{
Same as Fig. 11, but for models A2 and A3.
\label{fig12}}
\end{figure}

Calculated velocity distributions of $\rm ^{56}Ni$ are shown in
Figures~\ref{fig11} and~\ref{fig12}. Velocity distributions are
calculated assuming that the angle between the line of sight and the
symmetry axis is $44^{\circ}$,  which is inferred from the form of the
ring around SN 1987A (\cite{plait95}). Asymmetric forms of the mass cut 
(A7) are adopted for the jet-like explosion models. As can be seen from
Figure~\ref{fig11}, fast moving component can not be reproduced in the 
spherical explosion models, which is consistent with other works
(\cite{herant91}; \cite{herant92}). On the other hand, fast moving
component is reproduced in the jet-like explosion models when the
amplitude of the initial fluctuations is set to be 30$\%$. However,
the slow moving component becomes insufficient as $\alpha$
gets larger. In particular, the jet-like
explosion models with no perturbation suffer from this problem.

\placefigure{fig11}
\placefigure{fig12}

\begin{figure}
\epsscale{1.0}
\plotone{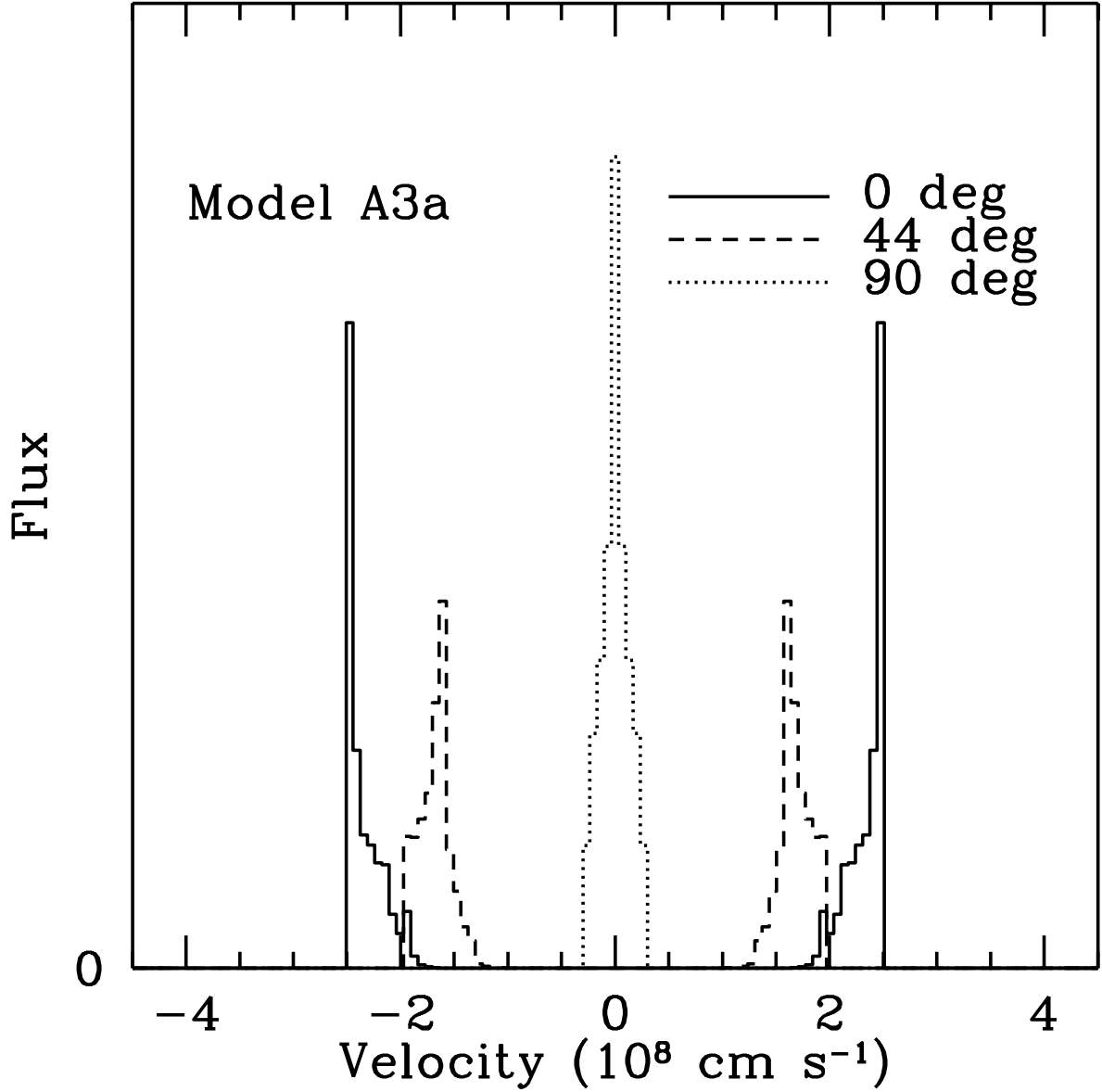}
\figcaption{
Velocity distributions for the model A3a seen from $\theta =
0^{\circ}$ (solid curve), $44^{\circ}$ (short-dashed curve), and
$90^{\circ}$ (dotted curve).
\label{fig13}}
\end{figure}

In order to understand the cause of the problem, velocity
distributions of A3a seen from $\theta = 0^{\circ}, 44^{\circ},$ and
$90^{\circ}$ are calculated. The result is shown in Figure~\ref{fig13}.
When the line of sight is parallel to the polar axis ($\theta =
0^{\circ}$), the matter that contains much of $\rm ^{56}Ni$ is
travelling at a maximum speed ($\sim 2500$ km/s). On the other hand,
when the system is seen from the equatorial axis, the apparent
velocities of the matter are almost zero. This proves that the matter
which contains much of $\rm ^{56}Ni$ is travelling along with the
polar axis. This picture can be easily explained as follows.
As can be seen in Figure~\ref{fig3}, $\rm ^{56}Ni$ can be synthesized
only in the polar region in the model A3. The synthesized $\rm
^{56}Ni$ behind the jet-like shock wave is moving along with the polar
axis. Although this feature is diluted by the presence of the
fluctuations, which introduce the slow moving component, the jet-like 
explosion models suffer from this problem.

\placefigure{fig13}

\begin{figure}
\epsscale{1.0}
\plotone{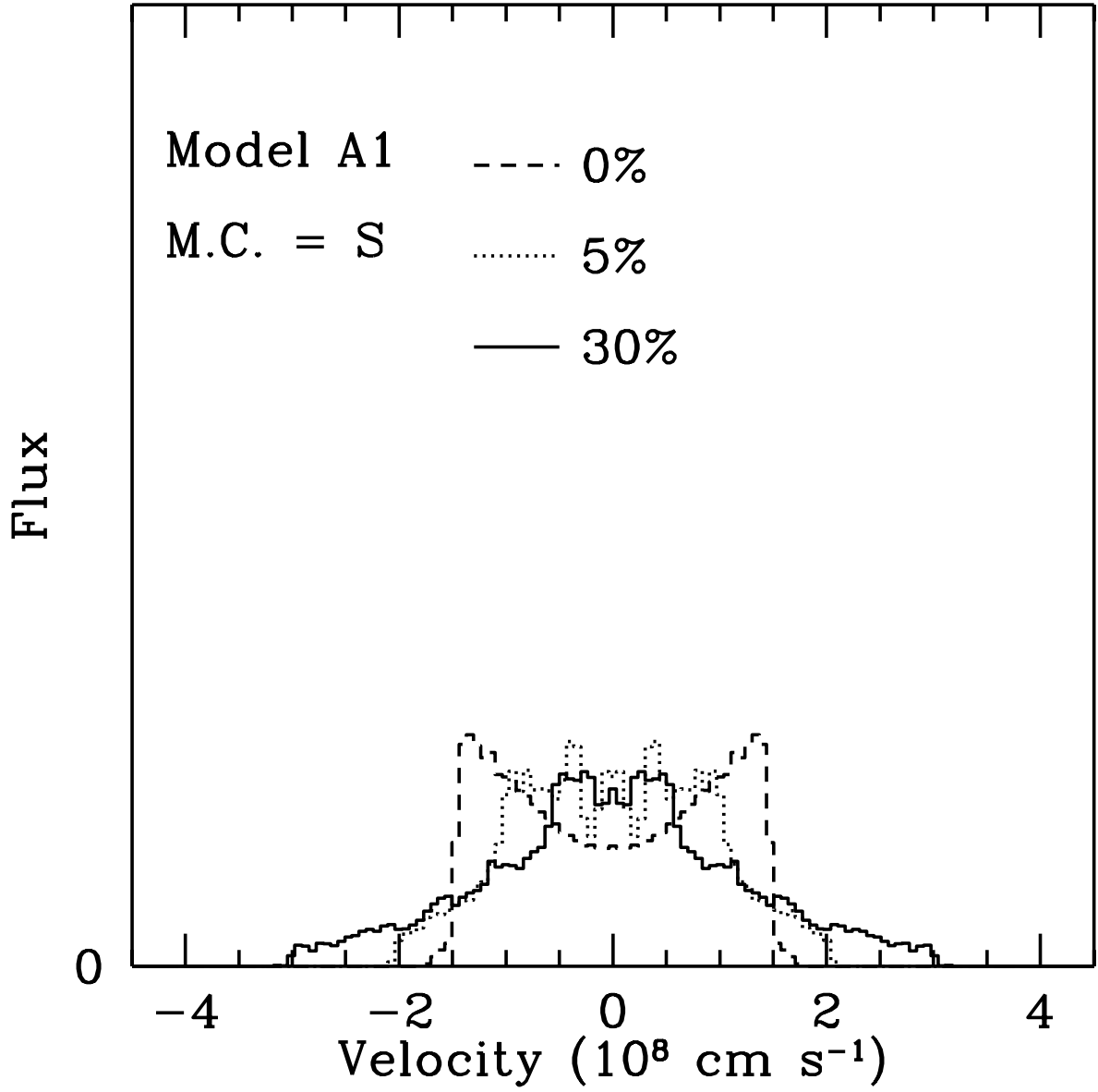}
\figcaption{
Velocity distributions for the model A1 with the spherical mass cut S7.
\label{fig14}}
\end{figure}

\begin{figure}
\epsscale{1.0}
\plottwo{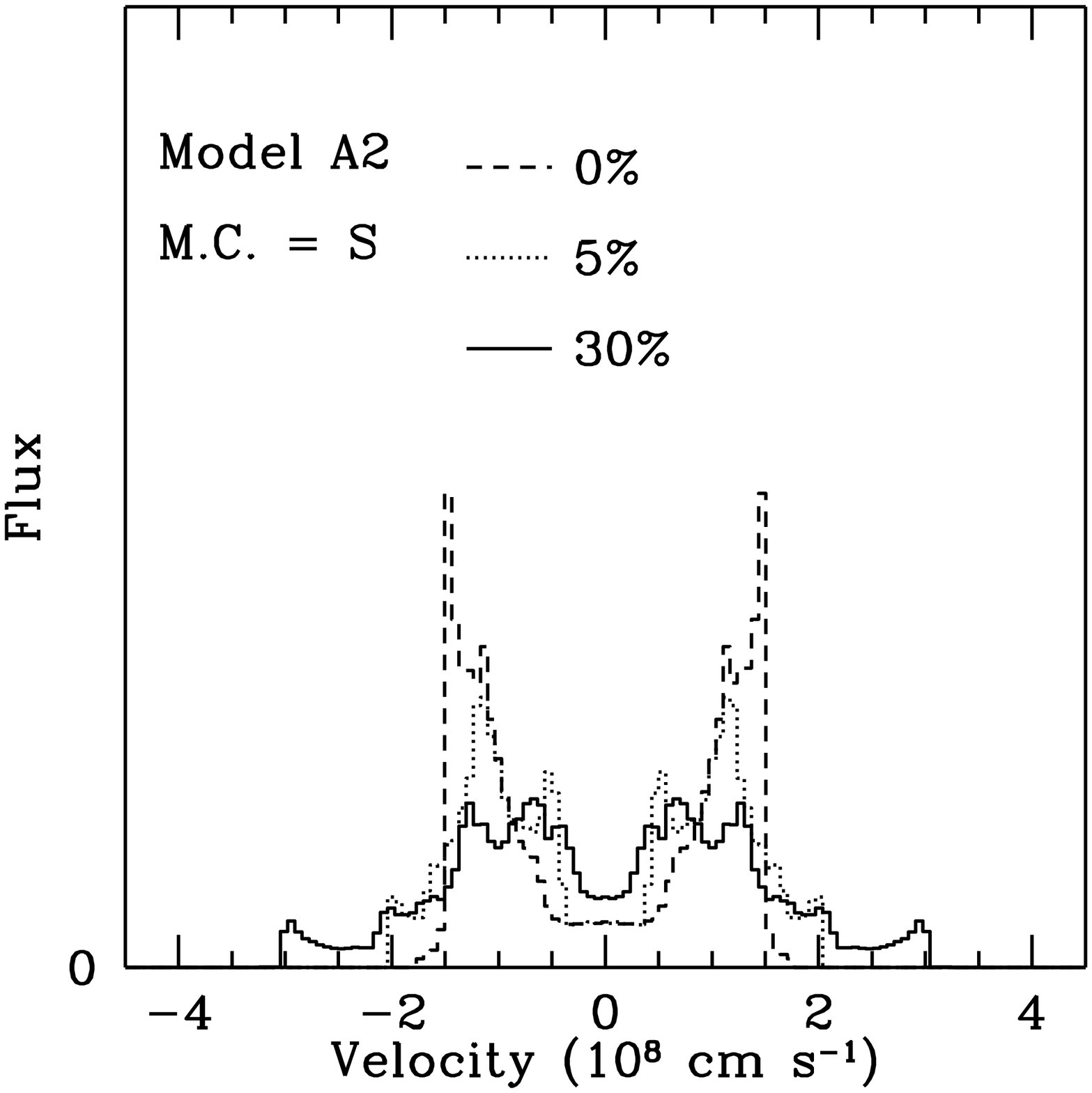}{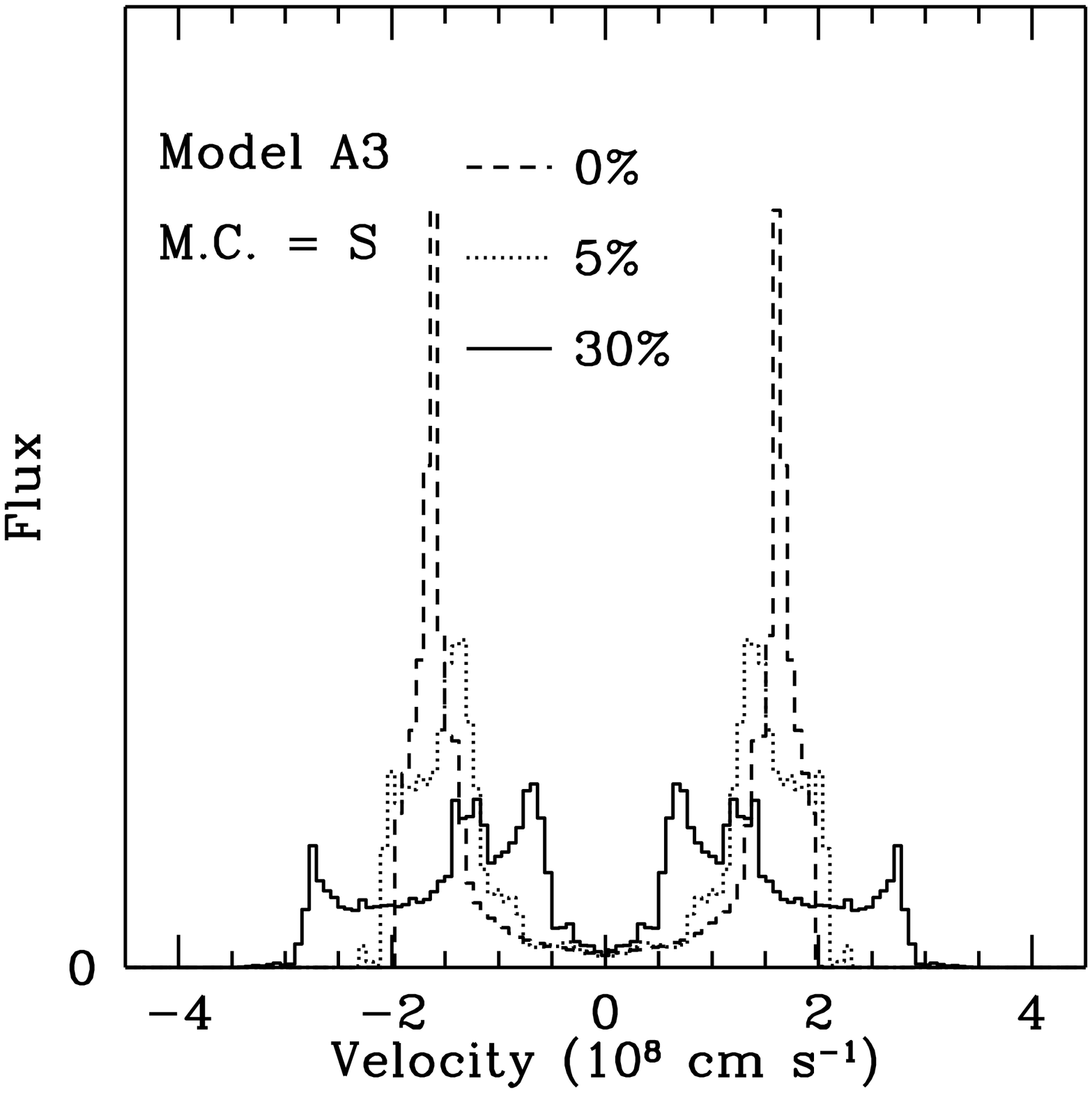}
\figcaption{
Same as Fig. 14, but for models A2 (left) and A3 (right).
\label{fig15}}
\end{figure}

\begin{figure}
\epsscale{1.0}
\plotone{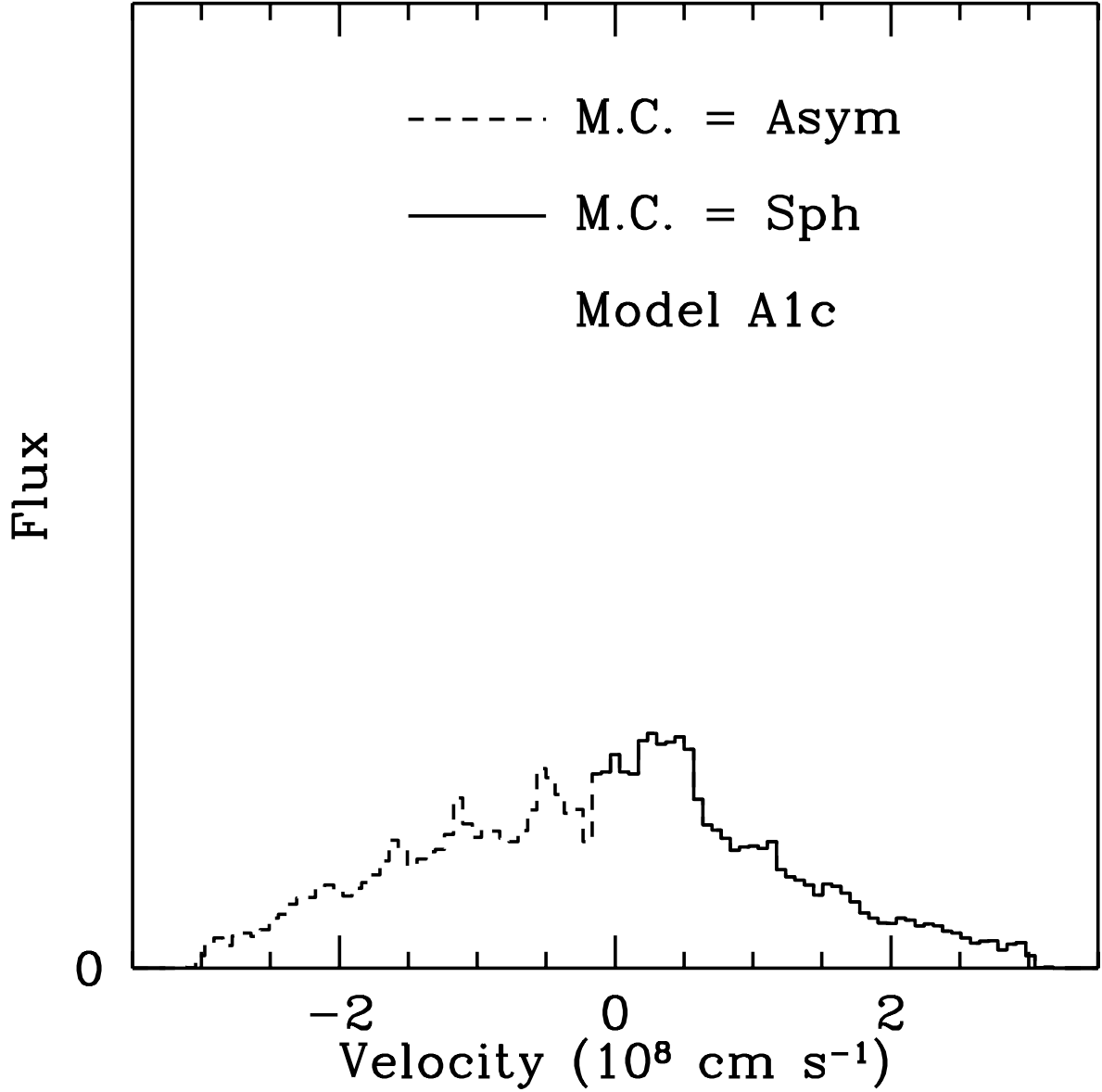}
\figcaption{
Combination of the model A1c with the spherical mass cut (S7) and
that with the asymmetric mass cut (A7).
\label{fig16}}
\end{figure}

Next, we investigate the dependence of the results on the form of the
mass cut. We can easily guess that a spherical mass cut will introduce 
the slow moving component around the equatorial plane, which may make
up for the deficiency. The results are presented in
Figures~\ref{fig14} and~\ref{fig15}. We can find the enhancement of
the slow moving component as we expected. However, the velocity
distributions of the models A2 and A3 are still far from the observed
line profiles. On the other hand, the velocity distribution of the
model A1c is similar to the observed one. Furthermore, we remove the
assumption of the equatorial symmetry of the system in order to
reproduce the asymmetry of the observed line profiles. We found that
the combination of the model A1c with the spherical mass cut (S7) and
that with the asymmetric mass cut (A7) is the best one for the
reproduction of the observed line profiles. This result is shown in
Figure~\ref{fig16}. This is an interesting result because it suggests
the origin of a kick velocity of a pulsar (\cite{gunn70}; \cite{lyne94}).
This is discussed in the next subsection.

\placefigure{fig14}
\placefigure{fig15}
\placefigure{fig16}

\begin{figure}
\epsscale{1.0}
\plotone{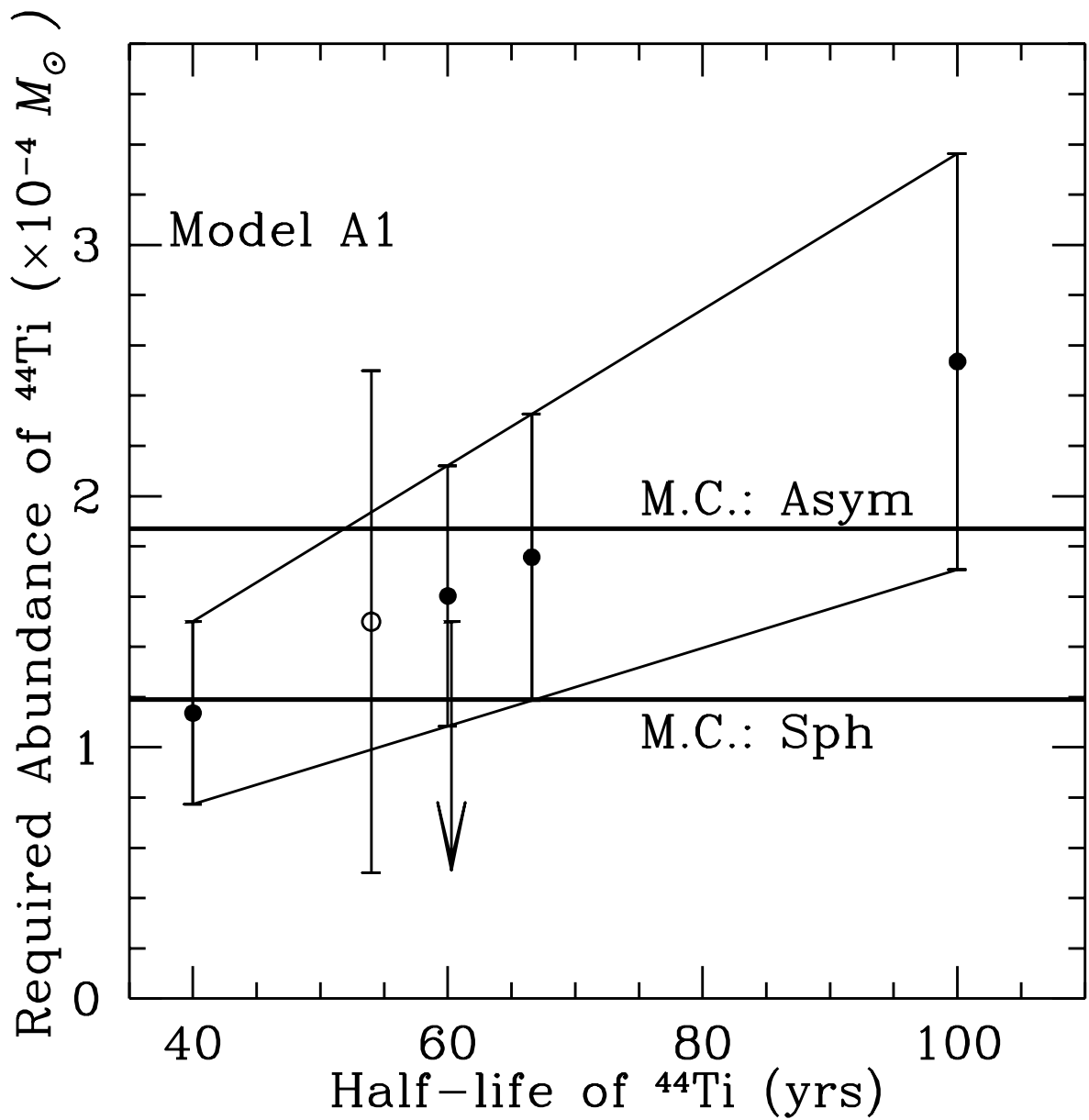}
\figcaption{
Horizontal lines: calculated amounts of $\rm ^{44}Ti$ in the model A1 with the
spherical/asymmetric mass cuts. Required amounts of $\rm ^{44}Ti$ are
also shown as a function of its half-life (Mochizuki \& Kumagai
1998; Mochizuki et al. 1999a; Kozma 1999; Lundqvist et al. 1999).
The most reliable value for its half-life is $\sim$ 60 yrs (Ahmad et
al. 1998; G$\rm \ddot{o}$rres et al. 1998; Norman et al. 1998).
\label{fig17}}
\end{figure}

Finally, we verify the consistency of the conclusions on the degree of 
the jet-like explosion in SN 1987A, which are derived by the calculations of
explosive nucleosynthesis and matter mixing.
We show again the required amounts of $\rm ^{44}Ti$ from the observed
luminosity evolution are shown in Figure~\ref{fig17}.
We show the calculated amounts of $\rm ^{44}Ti$ in the model A1 with the
spherical/asymmetric mass cuts in the figure. We can
find that the favored degree of a jet-like explosion to explain the
tail of the light curve is consistent with the one favored in the
calculation of the matter mixing. It is also noted that the
combination of the model A1 with S7 and that with A7 is also the best 
model that reproduces the observed luminosity of SN 1987A at $\sim$
3600 days after the explosion.

\placefigure{fig17}

\subsection{Discussion on the Matter Mixing} \label{discussion2}
\indent

A lot of discussions have already been presented in the previous
subsections. Additional comments are presented in this subsection.
We first discuss the form of the mass cut. As stated in
subsection~\ref{masscut}, the mass cut is determined by a guidance
principle. We also concluded that the results of the calculations on
the explosive nucleosynthesis, especially on the synthesis of $\rm
^{44}Ti$, are not sensitive to its form. However, we have to conclude
that the results of the calculations on the matter mixing are
sensitive. We conclude that a more spherical form than that determined
by the guidance principle is required by the calculations of the
matter mixing.

Next, we discuss the origin of a kick velocity of a pulsar. As shown
in Figure~\ref{fig16}, the observed line profiles are well reproduced
when the assumption of the equatorial symmetry of the system is
removed and the models A1c with the mass cuts A7 and S7 are combined.
Such a condition will be realized when the strength of the jet
generated at the north pole is different from that generated at the
south pole. Some models that realize such a condition have been
proposed. For example, asymmetries in density or velocity produced 
during the stellar evolution (\cite{bazan94}; \cite{bazan97}) may
cause such an asymmetric explosion (\cite{burrows96};
\cite{burrows96b}) in which the
amounts of the explosion energy in the northern and southern
hemispheres are different from each other.
Such an asymmetric explosion can be also caused by the effects of the
asymmetric magnetic field topology on the neutrino
transportation in the iron core (\cite{lai98}). 
If we believe these theories, the
proto-neutron star and the center of gravity of the ejecta are moving in
opposite directions due to the conservation of the momentum.
In case of SN 1987A, the jet on the northern hemisphere has to be stronger
and eject more $\rm ^{56}Ni$ than that on the southern hemisphere in
order to reproduce the observed asymmetric line profiles.
This is because many of the observations on the ring of SN
1987A show that the northern part of the ring is nearer to us
(\cite{michael98}), which means that the northern part of the jet is
red-shifted. In this case, the proto-neutron star born in SN 1987A is
moving in the southern part of the remnant. 
This prediction is consistent with the one that given Fargion \& Salis 
(1995), who discussed the effect of the precessing $\gamma$ jets
on the formation of the twin rings around SN1987A.
In particular, if the
axial symmetry of the system is kept well, the neutron star will be
running along with the polar axis. Detection of the neutron star in SN 
1987A will make a strong constraint on the mechanisms of pulsar kick
and jet-like explosion. Also, it is noted that the reproduced
asymmetric line profile in this study is only the first step of the
estimation of the effects of the asymmetric explosion. There are some
points to be refined as mentioned below. 
First, calculations of the matter mixing and explosive nucleosynthesis
have to be done with the zenith angle ranged between $0^{\circ}$ and
$180^{\circ}$ so that the gradients of the physical quanta on the
equatorial plane can be
treated correctly (these are assumed to be zero in all of the
calculations in this study because the symmetry with respect to the
equatorial plane is assumed from the beginning). This treatment will be
important when the explosion energy on the northern hemisphere is
different from that on the southern hemisphere, as the theories
mentioned above predict. Second, it is noted that the same amounts of
$\rm ^{56}Ni$ are assumed to be ejected from both hemispheres in this
study, which is not self-evident in an asymmetric explosion.
If the observations investigated in this study can be reproduced
better by these calculations, we will be able to predict the position and
velocity of the pulsar born in SN 1987A more definitely using results
of such advanced calculations.

We comment on the jet-induced explosion of a collapsing star
(\cite{macfadyen98}; \cite{khokhlov99}).
It is a very attractive theory because it has a possibility to explain 
a lot of observational facts and to be an origin of the $\gamma$-ray
bursts (\cite{macfadyen98}; \cite{khokhlov99}).
When we believe the correlation between a $\gamma$-ray
burst and a super(hyper)nova (\cite{kulkarni98}; \cite{galama98}),
such a jet-induced explosion model will be a convincing one for the
$\gamma$-ray burst model. In this case, we note, a bipolar profile of
iron like the one seen in Figure~\ref{fig13} should be observed when a
$\gamma$-ray
burst occurs near us, that is, in our galaxy and the Magellanic clouds.
Such an observation will give a strong support to this model.
However, it has to be also noted that such a strong jet-induced
explosion is not allowed for the model of SN 1987A by the calculations
of the explosive nucleosynthesis and the matter mixing investigated in
this study. Here we introduce the recent work presented by Fryer \&
Heger (1999) who performed core-collapse simulations using a rotating
progenitor star (\cite{heger99}). It should be emphasized that they
calculated the evolution of a star from the beginning of the
main-sequence stage including the effect of rotation. This
means that the distribution of the angular momentum in the iron core in
their progenitor model is not an artificial input parameter but a naturally
derived outcome. As a result,
they report the ratio of the escaping velocity of the matter behind
the shock wave in the polar region relative to the one in the
equatorial plane to be about two in the iron core.
That is, their result is entirely consistent with our
conclusion. We think such a moderate 'jet-like' explosion model is a
good one for the model of SN 1987A.

\section{SUMMARY} \label{summary}
\indent

In this study, calculations of the explosive nucleosynthesis are
performed in order to investigate the effects of the jet-like
explosion in a collapse-driven supernova. We found that the
radioactive nuclei $\rm ^{44}Ti$ is produced in a sufficient amount to 
explain the observed luminosity in SN 1987A at 3600 days
after the explosion. This is because the active alpha-rich freezeout
takes place behind the strong shock wave in the polar region.
The favored degree of a jet-like explosion to explain the observed
luminosity is concluded to be $\alpha \sim 1/3$, which means the ratio
of the velocity
along to the polar axis relative to that in the equatorial plane at the Si/Fe
interface is $\sim 2:1$. Such a high entropy condition caused by a
jet-like explosion will be also good for the synthesis of the
rapid-process nuclei (\cite{woosley94}; \cite{mclaughlin96}) in the iron
core. It will be a good challenging to perform a calculation of the
rapid-process nucleosynthesis in the iron core in order to investigate
the effects of the jet-like explosion. Additionally, a hypernova model
(\cite{iwamoto98}; \cite{woosley99}) will be also able to realize
such a high entropy condition. We are performing such calculations.
Results will be presented in the near future.

Calculations of the matter mixing in SN 1987A are also performed.
We found that the fast moving component of $\rm ^{56}Ni$ can be 
reproduced in the jet-like explosion models as long as the initial
velocity fluctuations is set to be $30 \%$, which can not be
reproduced in the spherical explosion models. On the other hand, 
the slow moving component becomes insufficient as $\alpha$ gets
larger. This is because $\rm ^{56}Ni$ is synthesized only in the polar 
region and accelerated along to the polar direction in an extreme
jet-like explosion model.
The favored degree of a jet-like explosion to explain the observed
line profiles of $\rm Fe[II]$ is $\alpha \sim 1/3$, which is
consistent with the one favored in the calculation of the explosive
nucleosynthesis. The required amplitude ($\sim 30 \%$) for the initial 
velocity
fluctuations supports that the origin of the fluctuations is the dynamics 
of the core collapse rather than the convection in the progenitor.
The asymmetry of the observed line profiles can be explained
when the assumption of the equatorial symmetry of the system is
removed, which can be caused by the asymmetry of the jet-like
explosion with respect to the equatorial plane.
When we believe some theories that cause such an asymmetric explosion, the
proto-neutron star born in SN 1987A will be moving in the southern part of
the remnant. Moreover, when the neutron star
is found in the remnant of SN 1987A and its position angle is found to 
be aligned with the direction of elongation of the SN debris,
it strongly supports the fact that the axial symmetry
is kept in the system.

The concluded degree of the jet-like explosion will give good constraints 
on the calculations of the dynamics of the collapse-driven supernovae.
We have to search for the proper conditions of the iron core, such as the
angular momentum, the strength of the magnetic field, and the form
of the neutrino sphere, in order to generate the required
shock wave in this study. Such efforts will give a new insight to
the dynamics of the collapse-driven supernovae and the feature of the
proto-neutron stars.

\acknowledgements
I would like to thank Prof. A. Burrows and Prof. D. Fargion for
useful comments on the mechanism of a pulsar kick. 
I express my thanks to Prof. Lundqvist and Dr. C. Kozma for their kind
comments on the luminosity evolution in SN 1987A.
The author is also grateful to Prof. K. Sato and Dr. S. Yamada for
useful discussions. Finally, I would like to thank the anonymous
referee for his helpful comments. 
This research has been supported in part by a Grant-in-Aid for the
Center-of-Excellence (COE) Research (07CE2002) and for the Scientific
Research Fund (199908802) of the Ministry of Education, Science, Sports and
Culture in Japan and by Japan Society for the Promotion of Science
Postdoctoral Fellowships for Research Abroad.

\clearpage

\begin{table*}
\begin{center}
\begin{tabular}{ccccccccccccccccc}
\tableline
\tableline
      &           &  Fluctuation    &       &           &  Fluctuation\\
Model &  $\alpha$ &  ($\%$)         & Model & $\alpha$  &  ($\%$) \\
\tableline
S1a   &   0       &    0            & A2a   & 3/5       & 0       \\
S1b   &   0       &    5            & A2b   & 3/5       & 5       \\
S1c   &   0       &    30           & A2c   & 3/5       & 30      \\
A1a   &   1/3     &    0            & A3a   & 7/9       & 0       \\
A1b   &   1/3     &    5            & A3b   & 7/9       & 5       \\
A1c   &   1/3     &    30           & A3c   & 7/9       & 30      \\
\tableline
\end{tabular}
\tablenum{1}
\caption{
Models, Values of $\alpha$, and Amplitudes of the fluctuations. Since the
initial velocity fluctuation is not introduced in the calculations of
the explosive nucleosynthesis, we name the models S1, A1, A2, and A3 for
the calculations.
}\label{tab1}
\end{center}
\end{table*}

\begin{table*}
\begin{center}
\begin{tabular}{ccccccccccccccccc}
\tableline
\tableline
Element & $\rm A_{min}$ & $\rm A_{max}$  & Element & $\rm A_{min}$ &
$\rm A_{max}$ & Element & $\rm A_{min}$ & $\rm A_{max}$ \\
\tableline
N & 1 & 1 & Al& 24 & 30 & V & 44 & 54 \\
H & 1 & 1 & Si& 26 & 33 & Cr& 46 & 55 \\
He& 4 & 4 & P & 28 & 36 & Mn& 48 & 58 \\
C & 11& 14& S & 31 & 37 & Fe& 52 & 61 \\
N & 12& 15& Cl& 32 & 40 & Co& 54 & 64 \\
O & 14& 19& Ar& 35 & 45 & Ni& 56 & 65 \\
F & 17& 22& K & 36 & 48 & Cu& 58 & 68 \\
Ne& 18& 23& Ca& 39 & 49 & Zn& 60 & 71 \\
Na& 20& 26& Sc& 40 & 51 & Ga& 62 & 73 \\
Mg& 22& 27& Ti& 42 & 52 & Ge& 64 & 74 \\
\tableline
\end{tabular}
\tablenum{2}
\caption{
Nuclear Reaction Network Employed
}\label{tab2}
\end{center}
\end{table*}

\begin{table*}
\begin{center}
\begin{tabular}{ccccccccccccccccc}
\tableline
\tableline
  & Ha95 & WW95 & TNH96 & Na97 \\
\tableline
$M\rm (^{56}Ni)$ & 0.073 & 0.088 & 0.074 & 0.070 \\
$\langle \rm ^{57}Ni/^{56}Ni \rangle$ & 1.7 & 0.9  & 1.6 & 1.5 \\
$\langle \rm ^{58}Ni/^{56}Ni \rangle$ & 1.2 & 0.87 & 1.3 & 1.5 \\
$M \rm ( ^{44}Ti)$              & 0.915 & 0.138 & 1.53 & 0.646 \\
\tableline
\end{tabular}
\tablenum{3}
\caption{
Results of the numerical calculations for
the explosive nucleosynthesis in a collapse driven supernova assuming
a spherical explosion. Ha95, WW95, TNH96, and Na97 represent Hashimoto 
(1995), Woosley \& Weaver (1995), Thielemann, Nomoto, Hashimoto
(1996), and Nagataki et al. (1997), respectively. WW95 represents the
model S20A in their paper. Mass of the
progenitor is assumed to be $\sim$ 20$M_{\odot}$ in the main-sequence
stage and to have had $\sim$ 6$M_{\odot}$ helium core, as a model of
the progenitor of SN 1987A, Sk-69$^{\circ}$202. 
$M\rm (^{56}Ni)$ means the mass of $\rm ^{56}Ni$ in the ejecta, in
units of $M_{\odot}$.
The ratio $\langle \rm ^{57}Ni/^{56}Ni \rangle = \langle \rm
^{57}Co/^{56}Co \rangle$ is defined as $[\it X(\rm ^{57}Ni)/ \it X(\rm
^{56}Ni)]/[\it X(\rm ^{57}Fe)/\it X(\rm ^{56}Fe)]_{\odot}$.
$\langle \rm ^{58}Ni/^{56}Ni \rangle$ is defined as $[\it X(\rm
^{58}Ni)/ \it X(\rm ^{56}Ni)]/[\it X(\rm ^{58}Ni)/\it X(\rm
^{56}Fe)]_{\odot}$. $M \rm ( ^{44}Ti)$ is the ejected mass of $\rm
^{44}Ti$ in units of $10^{-4} M_{\odot}$.
}\label{tab3}
\end{center}
\end{table*}

\begin{table*}
\begin{center}
\begin{tabular}{ccccccccccccccccc}
\tableline
\tableline
Model & M.C & $\langle \rm ^{57}Ni/^{56}Ni \rangle$ & $\langle \rm
^{58}Ni/^{56}Ni \rangle$ & Mass of $\rm ^{44}Ti$ \\
\tableline
S1 & S7 & 1.5 & 1.5  & 6.46E-5 \\
A1 & S7 & 1.7 & 1.9  & 1.19E-4 \\
A2 & S7 & 1.7 & 1.8  & 1.61E-4 \\
A3 & S7 & 1.8 & 1.6  & 3.40E-4 \\
S1 & A7 & 1.5 & 1.5  & 6.46E-5 \\
A1 & A7 & 1.8 & 1.5  & 1.87E-4 \\
A2 & A7 & 1.8 & 1.3  & 2.97E-4 \\
A3 & A7 & 1.8 & 0.97 & 5.10E-4 \\
\tableline
\end{tabular}
\tablenum{4}
\caption{
Synthesized heavy elements in each model. M.C means the adopted form of the
mass cut. Mass of $\rm ^{44}Ti$ is written in units of $M_{\odot}$.
Value of $Y_e$ between $M = 1.637 M_{\odot}$ and the Si/Fe interface
is changed to that at $M = 1.637 M_{\odot}$ (= 0.499).
}\label{tab4}
\end{center}
\end{table*}

\end{document}